\documentclass[twocolumn]{aastex63}

\newcommand{\HII}{H\,{\sc ii}}
\newcommand{\NII}{[N\,{\sc ii}]}
\newcommand{\OIII}{[O\,{\sc iii}]}
\newcommand{\SII}{[S\,{\sc ii}]}

\newcommand{\Ha}{H$\alpha$}
\newcommand{\Hb}{H$\beta$}

\newcommand{\Msun}{~M$_{\odot}$}

\usepackage{natbib}
\usepackage{graphicx}	
\usepackage{amsmath}	
\usepackage{amssymb}	
\usepackage{bm}		
\usepackage{mathtools}
\usepackage{float}
\usepackage{xcolor}
\usepackage{latexsym}
\usepackage{amssymb}
\usepackage{longtable}
\usepackage{epsf}
\usepackage{enumitem}
\usepackage[none]{hyphenat}

\received{June 1, 2019}
\revised{January 10, 2019}
\accepted{\today}
\submitjournal{ApJ}

\shorttitle{On the ULX sources in NGC 925}
\shortauthors{Lara-L\'opez et al.}

\begin{document}

\title{Metal-THINGS: On the metallicity and ionization of ULX sources in NGC 925}

\correspondingauthor{Maritza A. Lara-L\'opez}
\email{Maritza.Lara-Lopez@armagh.ac.uk}

\author[0000-0001-7327-3489]{Maritza A. Lara-L\'opez}
\affiliation{DARK, Niels Bohr Institute, University of Copenhagen, Lyngbyvej 2, Copenhagen DK-2100, Denmark}
\affiliation{Armagh Observatory and Planetarium, College Hill, Armagh, BT61 DG, UK}

\author{Igor A. Zinchenko}
\affiliation{Faculty of Physics, Ludwig-Maximilians-Universit{\"a}t, Scheinerstr. 1, 81679 Munich, Germany}
\affiliation{Main Astronomical Observatory, National Academy of Sciences of Ukraine, 27 Akademika Zabolotnoho St, 03680 Kiev, Ukraine}

\author{Leonid S. Pilyugin}
\affiliation{Main Astronomical Observatory, National Academy of Sciences of Ukraine, 27 Akademika Zabolotnoho St, 03680 Kiev, Ukraine}
\affiliation{Astronomisches Rechen-Institut, Zentrum für Astronomie der Universität Heidelberg, Mönchhofstr. 12-14, 69120, Heidelberg, Germany}

\author{Madusha L. P. Gunawardhana}
\affiliation{Leiden Observatory, Leiden University, PO Box 9513, 2300 RA, Leiden, The Netherlands}

\author[0000-0002-1381-7437]{Omar L\'opez-Cruz}
\affiliation{Coordiaci\'{o}n de Astrof\'{i}sica, Instituto Nacional de Astrof\'{i}sica \'{O}ptica y Electr\'{o}nica (INAOE), Luis E. Erro No. 1, Sta. Ma. Tonantzintla, Puebla, C.P. 72840 M\'{e}xico}

\author{Shane P. O\textsc{\char13} Sullivan}
\affiliation{School of Physical Sciences and Centre for Astrophysics \& Relativity, Dublin City University, Glasnevin, D09 W6Y4, Ireland}

\author{Anna Feltre}
\affiliation{INAF - Osservatorio di Astrofisica e Scienza dello Spazio di Bologna, Via P. Gobetti 93/3, 40129 Bologna, Italy}

\author[0000-0003-1113-2140]{Margarita Rosado}
\affiliation{Instituto de Astronom\'{i}a, Universidad Nacional Autonoma de M\'{e}xico, Apartado Postal 70-264, CP 04510 M\'{e}xico, CDMX, M\'{e}xico}

\author{M\'onica S\'anchez-Cruces}
\affiliation{Aix Marseille Univ, CNRS, CNES, LAM, Laboratoire d'Astrophysique de Marseille, Marseille, France}

\author{Jacopo Chevallard}
\affiliation{Sorbonne Universit\'{e}, UPMC-CNRS, UMR7095, Institut d\textsc{\char13}Astrophysique de Paris, F-75014, Paris, France}

\author{Maria Emilia De Rossi}
\affiliation{Universidad de Buenos Aires, Facultad de Ciencias Exactas y Naturales y Ciclo B\'{a}sico Com\'{u}n. Buenos Aires, Argentina}
\affiliation{CONICET-Universidad de Buenos Aires, Instituto de Astronom\'{i}a y Física del Espacio (IAFE). Buenos Aires, Argentina}

\author{Sami Dib}
\affiliation{Laboratoire d'Astrophysique de Bordeaux, Universit\'{e} de Bordeaux, CNRS, B18N,  all\'{e}e Geoffroy Saint-Hilaire, 33615, Pessac, France}

\author{Jacopo Fritz}
\affiliation{Instituto de Radioastronom\'{i}a y Astrof\'{i}sica, UNAM, Campus Morelia, A.P. 3-72, C.P. 58089, M\'{e}xico}

\author{Isaura Fuentes-Carrera}
\affiliation{Escuela Superior de F\'{i}sica y Matematicas, Instituto Polit\'{e}cnico Nacional, U. P. Adolfo L\'{o}pez Mateos, Zacatenco, 07730 M\'{e}xico, D. F., M\'{e}xico}

\author{Luis E. Gardu\~{n}o}
\affiliation{Coordiaci\'{o}n de Astrof\'{i}sica, Instituto Nacional de Astrof\'{i}sica \'{O}ptica y Electr\'{o}nica (INAOE), Luis E. Erro No. 1, Sta. Ma. Tonantzintla, Puebla, C.P. 72840 M\'{e}xico}

\author{Eduardo Ibar}
\affiliation{Instituto de F\'isica y Astronom\'ia, Universidad de Valpara\'iso, Avda. Gran Breta\~na 1111, Valpara\'iso, Chile}






\begin{abstract}
{ We present an analysis of the optical properties of three Ultra Luminous X-ray (ULX) sources identified in NGC 925.  We use Integral field unit data from the George Mitchel spectrograph in the context of the Metal-THINGS survey. The optical properties for ULX-1 and ULX-3 are presented, while the spaxel associated with ULX-2 had a low S/N, which prevented its analysis.  We also report the kinematics and  dimensions of the optical nebula associated with each ULX using ancillary data from the PUMA Fabry-Perot spectrograph.
A BPT analysis demonstrates that most spaxels in NGC 925 are dominated by star-forming regions, including those associated with ULX-1 and ULX-3. Using the resolved gas-phase metallicities, a negative metallicity gradient is found, consistent with previous results for spiral galaxies, while the ionization parameter tends to increase radially throughout the galaxy.  
Interestingly, ULX-1 shows a very low gas metallicity for its galactocentric distance, identified by two independent methods, while exhibiting a typical ionization. We find that such low gas metallicity is best explained in the context of the high-mass X-ray binary population, where the low-metallicity environment favours active Roche lobe overflows that can drive much higher accretion rates. An alternative scenario invoking accretion of a low-mass galaxy is not supported by the data in this region. 
Finally, ULX-3 shows both a high metallicity and ionization parameter, which is consistent with the progenitor being a highly-accreting neutron star within an evolved stellar population region. }

\end{abstract}



\section{Introduction} \label{sec:intro}

Ultra Luminous X-ray (ULX) sources are intriguing phenomena as their exact origin is not yet clear. They are defined as extragalactic non-nuclear X-ray point sources with luminosities exceeding of 10$^{39}$ ergs s$^{-1}$. This is higher than the Eddington luminosity of a 10\Msun\  black hole, which raises the question of how ULXs attain such high luminosities.


The majority of ULXs are thought to be X-ray binaries powered by accretion onto a compact object (black hole or  neutron star) and their luminosities are comparable to or exceed the Eddington luminosity of stellar black holes (BHs). Thus, ULXs offer a means to study  accretion near and above the Eddington limit, if the compact objects are stellar mass black holes, or they otherwise represent the discovery of intermediate mass black holes (\citeauthor{Colbert99} \citeyear{Colbert99}, see \citeauthor{Kaaret17} \citeyear{Kaaret17} for a review). 



Most ULXs have been detected in galaxies with a high star formation rate (SFR), suggesting that these objects are associated with young stellar populations \citep{Irwin04,Soria05,Swartz08}. Indeed, several papers report  a correlation between  ULXs and low-metallicity environments \citep{Pakull02,Zampieri04, Soria05, Swartz08,Mapelli09,Mapelli10,Kaaret11}, suggesting that  a large fraction of ULXs can be powered by massive BHs formed by the collapse of massive metal-poor stars. 
{ Due to the different origin of ULXs, studies at multiple wavelengths can provide insights into the origin and environment of these objects. For instance, optical studies find several ULXs in the local Universe surrounded by large nebulae with sizes of a few hundred parsec, i.e., in NGC 1313, M81, and Holmberg II \citep{Pakull02, Wang02}.

{ Spectroscopic data can provide information about the photo- or shock-ionized nature of the nebula and the ULXs contribution. Photoionisation can be attributed to the high X-ray and UV luminosity of the ULX \citep[i.e.,][]{Pakull02, Soria06}, while shock-ionization to jets, outflows or disk winds \citep[i.e., ][]{Grise06, Cseh12}. Nonetheless, in several cases the nebula has both, photoionized and shock-ionized gas \citep[][]{Roberts03,Abolmasov07,Urquhart18,Lopez19}.}

Additionally, a correlation between SFR and the number of ULX sources  per galaxy, the SFR-N$_{\rm ULX}$ relation, is well established \citep{Grimm03,Gilfanov04,Ranalli03}. Furthermore, the frequency of ULXs, normalized to the galaxy's SFR, depends on metallicity.  This means that a metal-poor (Z $<$ 0.2 Z$_{\sun}$) galaxy tends to have more ULXs with respect to a relatively metal-rich (Z$ >$ 0.2 Z$_{\sun}$) galaxy with the same SFR \citep{Mapelli10}.}




{ This paper provides a complete analysis of the optical spectroscopic properties of three ULXs in NGC 925. With a SFR of 0.561 M$_{\odot} $yr$^{-1}$ \citep{Leroy08}, the prediction from the SFR-N$_{\rm ULX}$ scaling relation of \citet[][Eq. 3]{Mapelli10} for NGC 925 is 0.79 ULXs, implying NGC 925 hosts an unusually high number of these objects.}
The host galaxy NGC 925 { is at a distance of 9.2 Mpc \citep[][cepheid distance]{Saha06}}, and is classified as a late-type barred spiral with two arms. It was previously observed in HI by \citet{Pisano98}, and later by the THINGS survey \citep{Walter08}. NGC 925 is described in  \citet{Blok08} as having a weak bar. NGC 925 has a stellar mass of 10$^{9.9}$ M$_{\odot}$, and an HI mass of 10$^{9.8}$ M$_{\odot}$  \citep{Leroy08}. 

Up until 2018, two ULXs were reported in NGC 925: J022727+333443 (ULX-1) and J022721+333500 (ULX-2) \citep[see][]{Heida16, Lopez17, Pintore18}. \citet{Heida16} obtained long slit spectroscopy  of the two ULXs in NGC 925 using  Keck/MOSFIRE H-band. They find a clear correspondence with a RSG for ULX-2. As for ULX-1, they find  that the ionization source of the [Fe II] $\lambda$16440  and HI emission lines  in the spectra may be stronger than the typical shocks found in Galactic supernova remnants.

A later X-ray analysis with Chandra, XMM-Newton and NuSTAR by \citet{Pintore18}, showed that the properties of ULX-1 are typical of a super-Eddington accreting stellar-mass black hole with an X-ray luminosity of $\sim$ 2.5 $\times$ 10$^{40}$ ergs s$^{-1}$, and they classify it as a broadened disc ULX. They suggest that ULX-1 is seen at small inclination angles, possibly through the evacuated cone of a powerful wind originating in the accretion disc. The nature of ULX-2 is more uncertain, although they disfavor an intermediate black hole candidate.

{ Recently, a third ULX (ULX-3) was identified by \citet{Earnshaw20} in NGC 925. They identified ULX-3 in 2017 with Chandra, and analysis of archival data revealed that it was detected by Swift in 2011, as well as by XMM-Newton in  2017, although at a much lower luminosity, and it was not detected by Chandra in 2005. Thus, ULX-3 is not a single one-off transient event, but an object that undergoes repeated increases in flux.} The detections of ULX-3 so far appear to indicate an approximately bimodal flux distribution. According to \citet{Earnshaw20}, this variability could be explained by a large-amplitude superorbital period \citep[e.g.,][]{Brightman19}, or by an accreting neutron star that, on occasion, enters a propeller regime, in which the magnetospheric radius of the magnetic field is larger than the corotation radius of the accretion disc, creating a centrifugal barrier to mass accretion and causing the flux to drop correspondingly  \citep[e.g.,][]{Tsygankov16}.

This study is part of the Metal-THINGS survey, a novel multiwavelength survey that is obtaining IFU spectroscopy of The HI Nearby Galaxy Survey \citep[THINGS, ][]{Walter08}, to be described in detail in Lara-L\'{o}pez et al. (in preparation).  

{ To date, only a few studies have analyzed the optical properties of the gas around ULXs. Although it is known that ULXs are located in low metallicity environments, this hasn't been quantified with respect to the metallicity gradient of the host galaxy, which can be flat for galaxies with $\lesssim$ 10$^9$M$_{\odot}$, and decrease from the center to the outskirts in more massive galaxies, with slopes of up to $-0.4$ dex \citep[e.g.,][]{Pilyugin14,Belfiore17,Igor19a}. In this paper, we analyze the ISM properties of NGC 925, with an emphasis on the gas properties  around the location of each ULX with respect to their gradient within the galaxy.} { To complement our analysis and to determine the size of the nebulae around the ULXs, we make use of ancillary data from the high resolution Fabry-Perot interferometer PUMA \citep{Rosado1995}.}

This paper is structured as follows, in \S \ref{Observations} we present descriptions of the observations, data reduction and the estimation of the optical properties. In \S \ref{EmLineDiag} we analyze the resolved BPT diagrams for this galaxy.  A description of the gas metallicity and ionization is given in  \S \ref{MetIonGrad}. Finally, a discussion and conclusions are given in \S \ref{Sect:Discussion} and \S \ref{Conclusion}, respectively.


\section[]{Observations}\label{Observations}


In this paper we use two sets of observations for NGC 925; (i) IFU spectroscopy from the George Mitchel spectrograph to derive the optical properties, such as gas metallicity and ionization; and  (ii) complementary  \Ha\  data from the Fabry-Perot interferometer PUMA   to accurately measure the dimensions of the nebulae around the ULXs. The spatial resolution of PUMA is  $\sim$ 3 orders of magnitude higher than the IFU spectroscopy and hence complementary.

\subsection[]{IFU spectroscopy}

\begin{figure}[ht]
\begin{center}
{\includegraphics[width=0.5\textwidth]{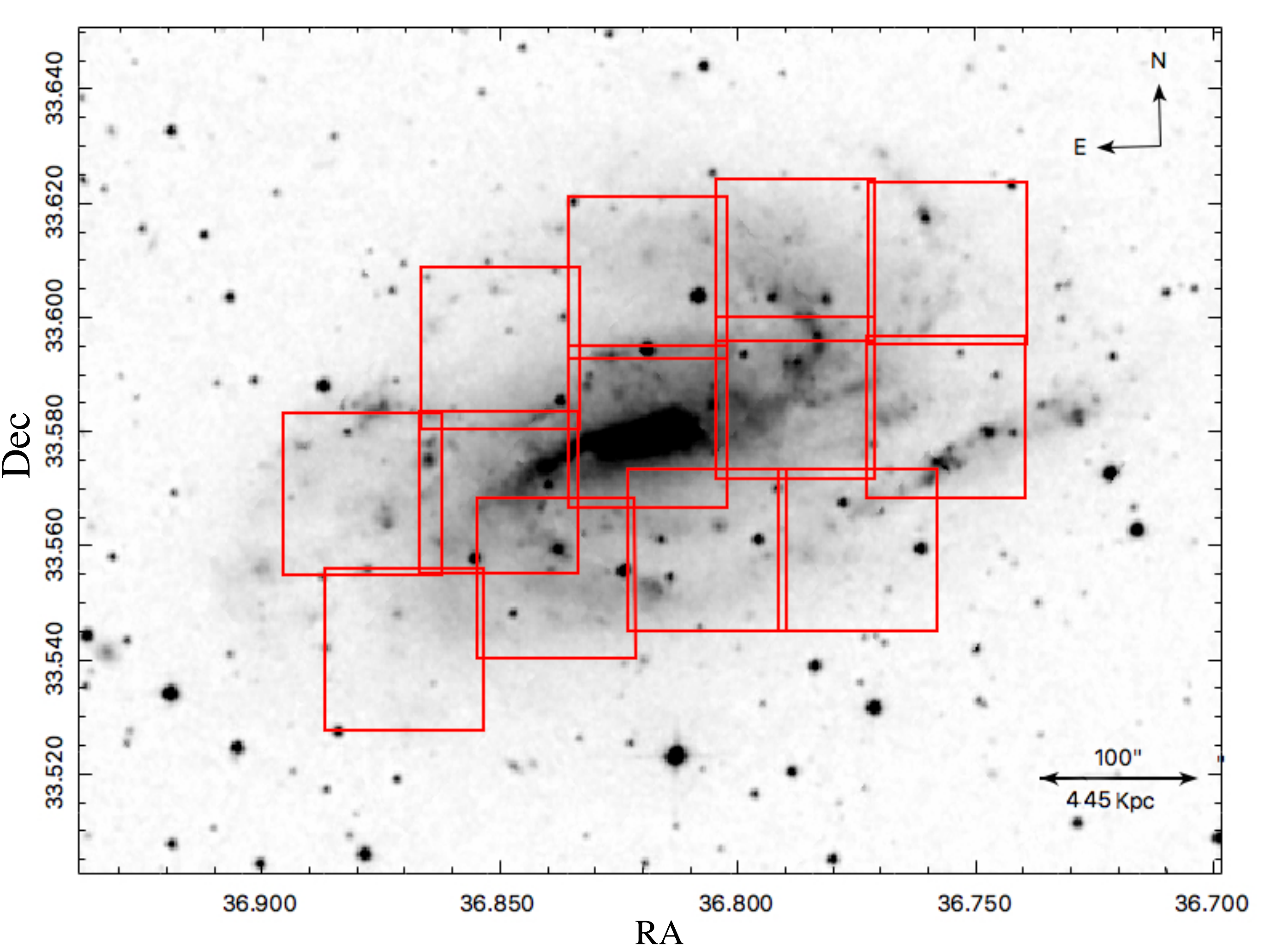}}
\end{center}
\caption{NGC 925 image showing the field of view of the GMS in red boxes, where a total of 13 pointings was observed. { (Archival image in B-band taken from the Palomar Observatory Sky Survey, NGS-POSS.}) \label{fig:Pointings}}
\end{figure}

\begin{figure*}[ht]
\begin{center}
{\includegraphics[width=0.9\textwidth]{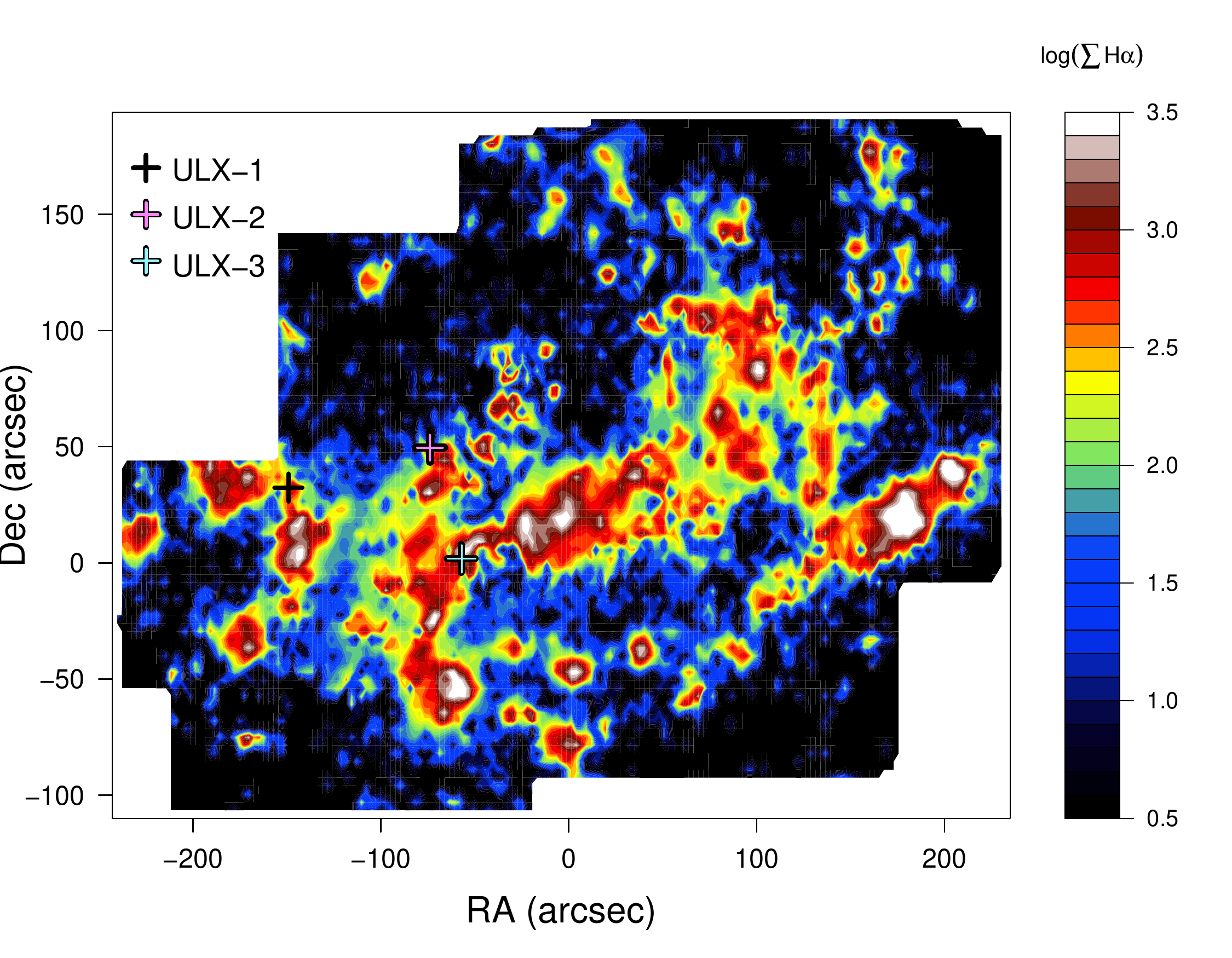}}
\end{center}
\caption{ H$\alpha$ surface brightness map of NGC 925, extinction corrected, in units of log(10$^{-17}$ ergs s$^{-1}$ cm$^{-2}$ arcsec$^{-2}$). Only spaxels with a S/N $>$ 3 in \Ha\ and \Hb\ are shown. The mosaic represents observations of 13 individual pointings using IFU data from GMS. The positions of ULX-1, ULX-2 and ULX-3 are shown with black, pink, and cyan crosses, respectively. As shown in Fig. \ref{fig:Pointings}, 100" correspond to 4.45 Kpc.}  \label{fig:HaMap}
\end{figure*}

The Metal-THINGS survey is a large program focused on obtaining  IFU spectroscopy for a unique sample of nearby galaxies  using the 2.7m Harlan Schmidt telescope at McDonald Observatory. It employs the George Mitchel spectrograph \citep[GMS, formerly known as VIRUS-P,][]{Hill08}, as well as the Multi Unit Spectroscopic Explorer \citep[MUSE, ][]{Bacon10} - on the Very Large Telescope (VLT). The selected galaxies are based on the THINGS survey \citep{Walter08}, which observed 34 large nearby  galaxies with the Very Large Array (VLA), obtaining HI data at high spatial and spectral resolution.

The GMS is a square array of 100 $\times$ 102 arcsecs, with a spatial sampling of 4.2 arcsecs, and a 0.3 filling factor. The IFU consists of 246 fibers arranged in a fixed pattern. NGC 925 was observed as part of the Metal-THINGS survey (Lara-Lopez et al. in preparation) with a red setup covering wavelengths from 4400 to 6800 \AA,  using the low resolution grating VP1, with a resolution of 5.3 \AA.  This wavelength range allows us to measure the following strongest emission lines:  \Hb, [OIII]$\lambda$$\lambda$4959, 5007, [OI]$\lambda$6300, \Ha, [NII]$\lambda$6584, and [SII]$\lambda$$\lambda$6716, 6731. 

Every pointing is observed with 3 dither positions to ensure a 90$\%$ surface coverage. Due to the extended nature of our galaxies, sky exposures are taken off-source for sky subtraction purposes during the reduction process. Our observing procedure consists of taking a 15 min exposure per dither, followed by a sky exposure, and the repetition of this process until 45 mins are reached per dither, per pointing. A calibration star was observed every night using 6 dither positions to ensure a 99$\%$ flux coverage. Calibration lamps (Neon $+$ Argon for the red setup) were observed at the beginning and end of every night to wavelength calibrate the spectra. The average seeing during the observations was 1.5 arcsecs. 

NGC 925 was observed in December 2017, January 2018, and January 2019 using GMS. A total of 13 pointings were observed (Fig \ref{fig:Pointings}),  which provide a total of 9,594 individual spectra for this galaxy. From our spectroscopic data, we estimate that NGC 925 is at a redshift of 0.00185, and adopted a cepheids-based distance of 9.2 Mpc \citep{Saha06}, which gives a scale of 0.044 kpc/", meaning each 4.2" fiber in the IFU corresponds to $\sim$184 pc.

The basic data reduction including bias subtraction, flat frame correction and wavelength calibration, was performed using P3D\footnote{https://p3d.sourceforge.io}. Sky subtraction and dither combination was performed using our own routines in Python. Flux calibration was performed following \citet{Cairos12} using 6 dither positions, ensuring a 97$\%$ coverage. We used IRAF \citep{IRAF1} to create a calibration function using the packages \textit{standard} and \textit{sensfunc}.

The stellar continuum of all flux-calibrated spectra was fitted using STARLIGHT \citep{Cid05, Mateus06, Asari07}. Briefly,  we used 45  simple stellar populations (SSP) models from the evolutionary synthesis models of \citet{Bruzual03} with ages from 1 Myr up to 13 Gyr and metallicities Z=0.005, 0.02 and 0.05. The stellar continuum  was subtracted from the spectra, and the emission lines were measured using Gaussian line-profile fittings. For a more detailed description, see \citet{Igor16, Igor19}.

The measured emission line fluxes were corrected for interstellar reddening using the theoretical H$\alpha$/H$\beta$ ratio and the reddening
function from \citet{Cardelli89} for $R_{V}$ = 3.1. 


Fig. \ref{fig:HaMap} shows the extinction corrected \Ha\  surface brightness map  for NGC 925. For visualization purposes, and only for this figure, an interpolation of the 13 pointings is displayed. In the same figure, the three identified ULXs  are marked with crosses. All further figures that use GMS data (Figs. \ref{fig:BPTMap} to \ref{fig:MetGrad}) show the data of individual fibers.

\subsection[]{PUMA Fabry-Perot observations}

To measure the size of the nebulae associated with the ULX sources, we use high spectral  and spatial resolution data from the  Fabry-Perot interferometer PUMA \citep{Rosado1995}. Observations of NGC 925 were done in December 2, 2018 at the 2.1 m telescope of the Observatorio Astron\'omico Nacional in San Pedro M\'artir, Baja California, M\'exico (OAN-SPM). We used a 2048 $\times$ 2048 spectral CCD detector, with a binning factor of 4, resulting in a field of view of 10$\arcmin$ in 512 $\times$ 512 pixels with a spatial sampling of 1.29$\arcsec$/pixel. Its Finesse ($\sim$24) leads to a sampling spectral resolution of 0.41 \AA\ (19.0 km s$^{-1}$). The spatial resolution of PUMA is 57.5 pc per pixel.


To isolate the redshifted H$\alpha$ emission of the galaxy, we used a filter centred at H$\alpha$ ($\lambda$6570 \AA)  with a full width at half maximum (FWHM) of 20 \AA. We scanned the  Fabry-Perot interferometer through 48 channels with an integration time of 120 s per channel, obtaining an object data cube of dimensions 512 $\times$ 512 $\times$ 48. 
The calibration data cube was obtained under the same conditions as the object data cube, just after its completion, and it has the same dimensions as the data cube. We used a Hydrogen lamp whose 6562.78 \AA\  line was close to the redshifted H$\alpha$ from the galaxy due to its low recession velocity.  The wavelength calibration of the PUMA object cube was carried out using the Fabry-Perot specialized software ADHOCw developed by J. Boulesteix \citep{Amram89} which uses the calibration cube to compute the phase map, a 2D map producing, pixel per pixel, the channel at which the H$\alpha$ line is at rest. Once wavelength calibrated, we obtained a lambda cube with 48 channels of different consecutive velocities. We also constructed a spatial smoothing (x,y) with a FWHM of (2,2) pixels. It is worth nothing that the data are not flux calibrated. 

The radial velocity field was obtained by computing the velocity value of the barycenter, the `center of mass' of the velocity profile, of each of the 512 $\times$ 512 line profiles as well as the gas velocity dispersion of the line profiles. These quantities were also obtained with the ADHOCw software.

Lastly, the velocity dispersion, $\sigma$, for each pixel was computed from the H$\alpha$ velocity profiles after correction from instrumental and thermal widths, assuming that all profiles (instrumental, thermal) are described by Gaussian functions. The velocity dispersion was estimated using $\sigma$ =
($\sigma^2_{\rm obs}$ - $\sigma^2_{\rm inst}$ - $\sigma^2_{\rm th}$)$^{1/2}$. Where, $\sigma_{\rm inst}$ = FSR/($\mathcal{F}$ $\times$ 2$\sqrt{2 {\ln} 2}$) = 16.4 km s$^{-1}$ and $\sigma_{\rm th}$ = 9.1 km s$^{-1}$. FSR = 912 km $^{-1}$ is the free spectral range of PUMA and $\mathcal{F}$ is its Finesse.
The thermal broadening was computed according to $\sigma_{\rm th}$ = (kT$_e$ /m$_{\rm H}$)$^{1/2}$ assuming an electronic temperature of T$_{\rm e}$ = 10$^{4}$ K \citep{Spitzer1978}.

\begin{figure*}[t]
{\includegraphics[width=1.1\textwidth]{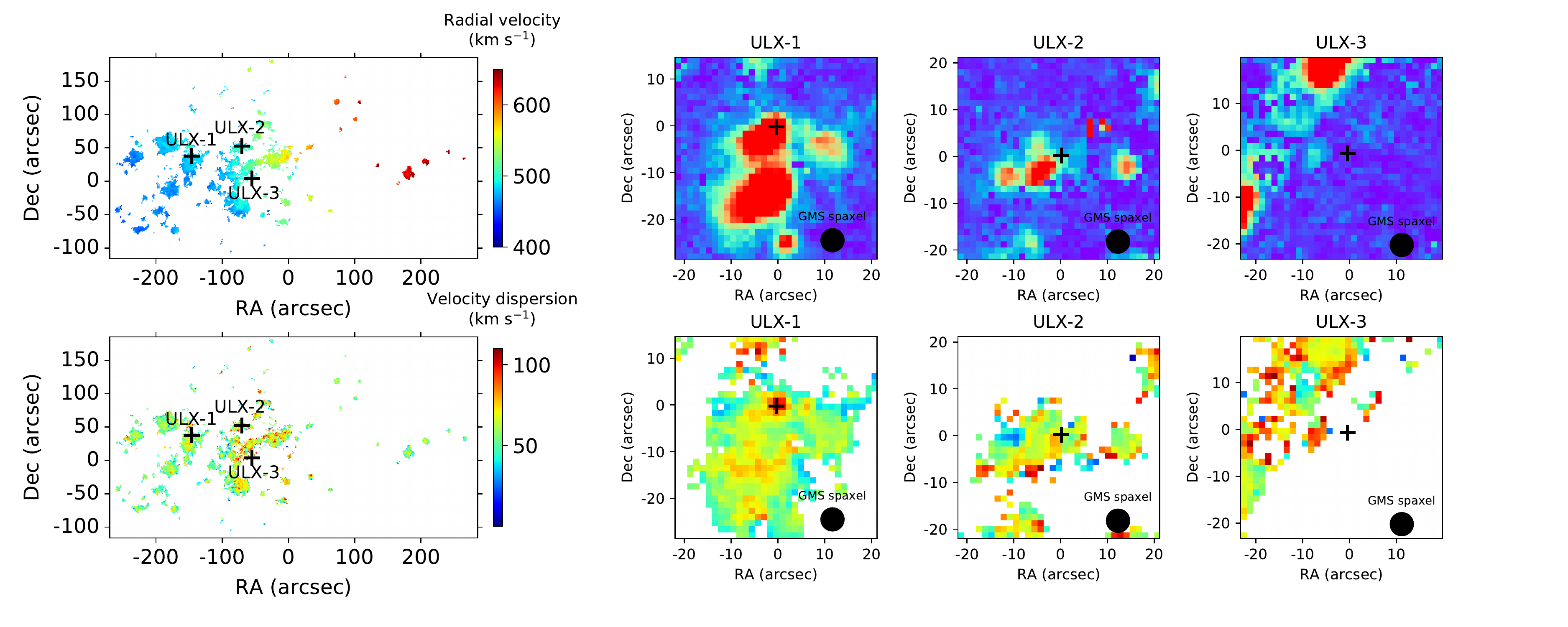}}
\caption{{ Left column: radial and velocity dispersion fields of NGC 925 obtained using the Fabry-Perot interferometer PUMA. The positions of the three ULX sources are marked with black crosses. Top right panels: close up of the H$\alpha$ high resolution monochromatic image of the three ULX sources. Bottom right panels: close up of the radial velocity field of the three ULX sources. The black filled circle represent a single GMS spaxel with 4.2$\arcsec$ of diameter.} }\label{fig:PUMA}
\end{figure*}

Figure \ref{fig:PUMA} depicts the main kinematic properties of the nebulae possibly associated with the ULX sources. The left column shows the radial velocity (top) and velocity dispersion (bottom) fields of NGC 925 obtained from the PUMA lambda cube. The right panels show close-ups of the monochromatic H$\alpha$ image { (not calibrated in flux)} and the radial velocity field, centred at the locations of the three ULXs.

We find that ULX-1 and ULX-2 are associated with large diameter nebulae. The dimensions of the nebulae are listed in Table  \ref{TableYY}. The diameters were measured from our monochromatic images taking into account the PUMA plate scale of 1.29$\arcsec$/pixel and assuming a distance to NGC 925 of { 9.2 Mpc}. The diameters of the nebulae  are similar to others found for ULXs located inside nebulae \citep[e.g.,][]{Pakull02, Abolmasov07}.


We integrated the radial velocity profiles of the pixels corresponding to each of the nebulae in order to get a global velocity profile of each nebula. We do not detect internal differences in the line profiles, at the limit of PUMA's velocity sampling resolution ($\sim$20 km s$^{-1}$). The systemic velocities of the nebulae follow the rotation of the galaxy, and are also listed in Table \ref{TableYY}. For ULX-3, we did not detect a nebulae, but two faint intensity blobs around the E-W direction separated by $\sim$350 pc. The size and distance to each blob is listed in Table \ref{TableYY}.


The range of velocity dispersion values in the nebulae associated with ULX-1 and ULX-2 are shown in Table 1, with values from 70-100 km s$^{-1}$ and 50-107 km s$^{-1}$, respectively. The velocity dispersion of the blobs associated with ULX-3 range from 26-104 km s$^{-1}$ for E$_{\rm blob}$ and  47-100 km s$^{-1}$ for W$_{\rm blob}$.  The high velocity dispersion of the nebulae and blobs associated with the ULXs  suggest the presence of shocks ionizing the gas. However, we cannot resolve any expansion signature,  presumably driven by an expanding bubble (ULX bubble) nor jets from the ULX source. On the other hand, ULX bubbles can be distinguished from supernova remnants (SNRs) and HII regions due to their larger size \citep[diameters of $\sim$100-300 pc, e.g.,][]{Pakull08}, comparable to the ones found for ULX-1 and ULX-2 in this work.

\begin{table*}
    	\caption{Dimensions and velocities of the ionized regions surrounding each of the ULX sources in NGC 925. { The distance to the ULX-Nebula is measured from the nebula's center to the ULX center.}}
	\begin{center}

    \hspace{-1cm} \scalebox{0.9}{
    \begin{tabular}{lccccc}
		\hline
 { } &
{ ULX-1} &
{ ULX-2} &
\multicolumn{2}{c}{{ ULX-3}}\\
\hline
{ Coordinates (J2000)} & {$\alpha$ = 02$^{h}$ 27$^{m}$ 27$^{s}$} & {$\alpha$ = 02$^{h}$ 27$^{m}$ 21$^{s}$} & \multicolumn{2}{c}{{$\alpha$ = 02$^{h}$ 27$^{m}$ 20$^{s}$.18}} \\
  & {$\delta$ = +33$^{\circ}$ 34$\arcmin$ 43$\arcsec$} &  {$\delta$ = +33$^{\circ}$ 35$\arcmin$ 00$\arcsec$} & \multicolumn{2}{c}{{$\delta$ = +33$^{\circ}$ 34$\arcmin$ 12$\arcsec$.84}}\\
 \cline{4-5}
 & & & {E$_{\rm blob}$} & {W$_{\rm blob}$}\\ 
 \cline{4-5}
{ Axis size major/minor (pc)}	& {285/285} 	& {285/232}  & {174/125} 	& {218/178} \\ 
{ Distance to ULX-Nebula (pc)}	& {143} & {156} &{172}  &{172} \\
{ Heliocentric velocity (km s$^{-1}$)}	&{ +495}  &{ +508} & {+495}  &{+525}  \\
{ Velocity dispersion (km s$^{-1}$)}& {70-100} & {50-107} &{26-104} & {47-100}\\  
 \hline
\end{tabular}} \\
\label{TableYY}
	\end{center}
\end{table*}

%
%

\section[]{Emission line diagnostic diagrams} \label{EmLineDiag}

\begin{figure*}[t]
{\includegraphics[width=0.5\textwidth]{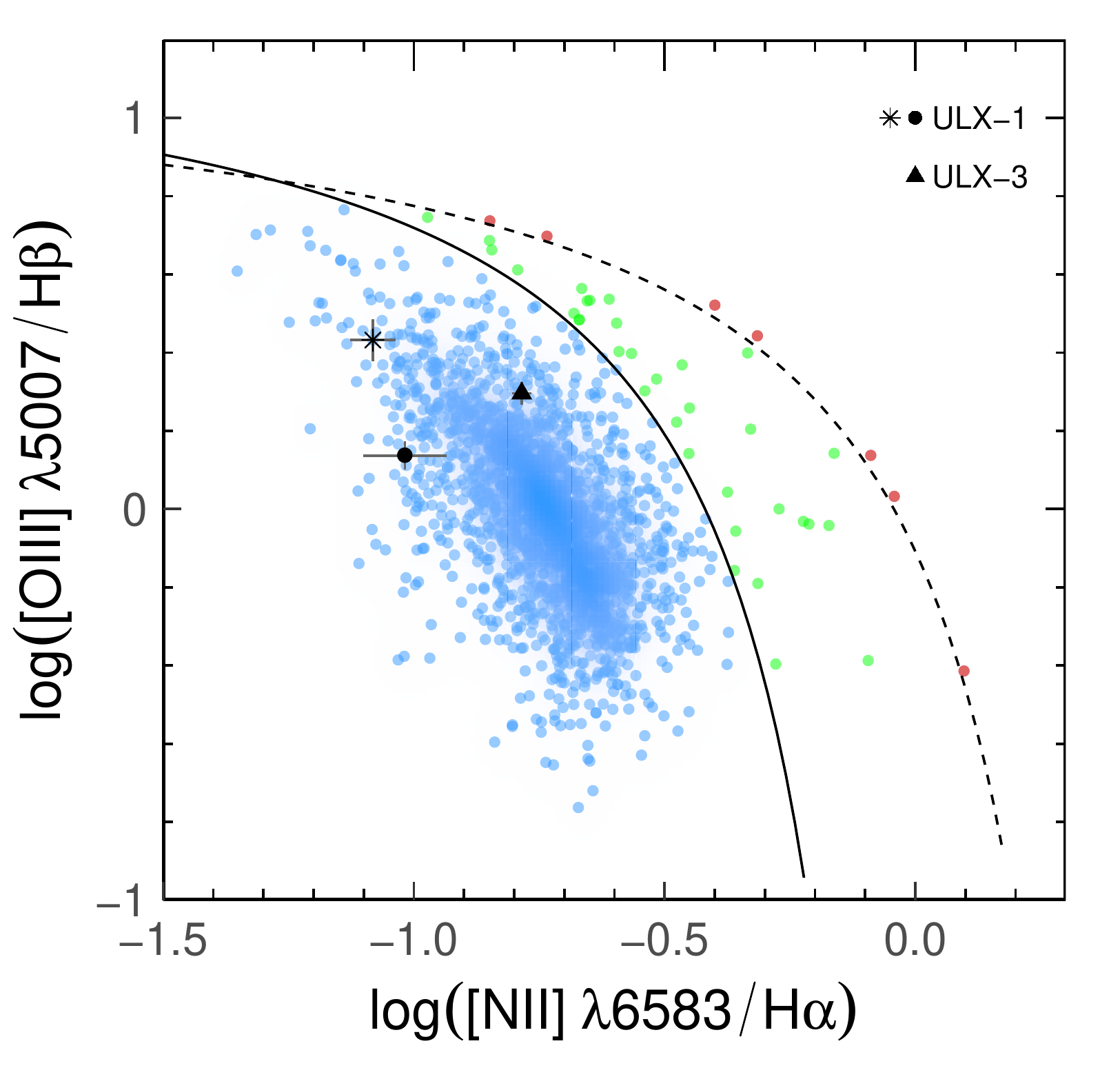}}
{\includegraphics[width=0.5\textwidth]{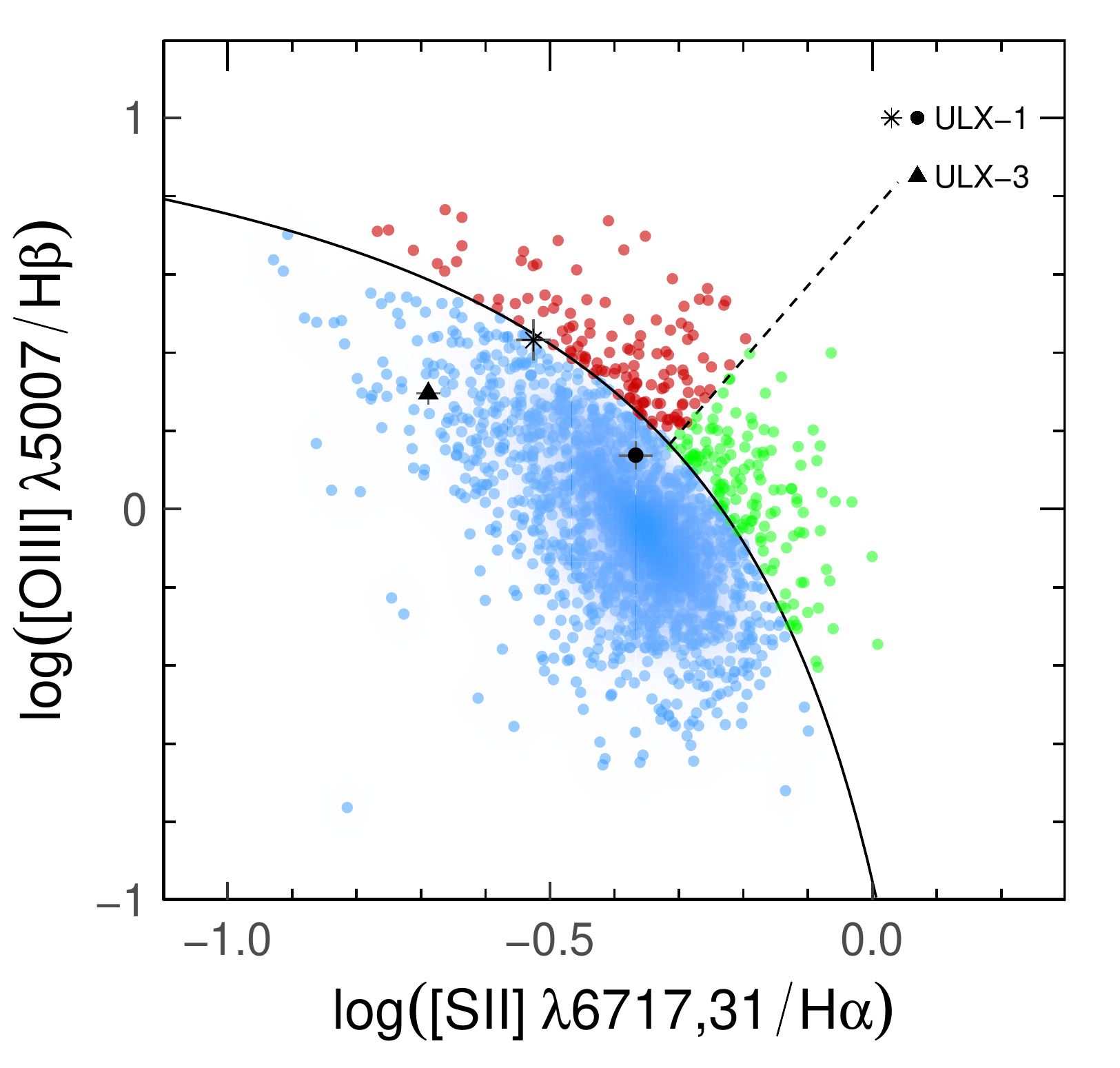}}
\caption{The left panel shows the \NII-BPT diagram with SF-type regions in blue, according to  K03 (black solid line). Green and red circles show spaxels classified as Composite and shocks/AGN types, respectively, using the criteria of \citet{Kewley06}. The right panel shows the  \SII-BPT diagram with SF-type regions in blue selected using the classification of  K03 (black solid line). Green and red circles show galaxies classified as LIER \citep{Belfiore16} and Seyfert-type regions \citep{Kewley06}, respectively. ULX-1(\textbullet , $\ast$) and ULX-3($\blacktriangle$) are marked in black symbols, { with their respective error bars in grey.}} \label{fig:BPTMap}
\end{figure*}

%
%


ULXs can have a strong effect on their surrounding nebula, either photoionizing or shock ionizing them. 
Line diagnostic diagrams provide valuable constraints on the nature of the ionizing source. If the nebula is photo-ionized, it is possible to measure its metallicity, while if it's shock-ionized, the mechanical power of the shock can be measured.


In this section, we analyze resolved diagnostic diagrams \citep[BPT diagrams,][]{Baldwin81, Veilleux87}, to identify regions with different types of ionization. 
Since these diagrams require several emission lines, we use the IFU data from the individual fibers of the GMS. Only spaxels with a S/N $>$ 3 for each of the used emission lines are taken into account.
In Fig. \ref{fig:BPTMap} (left),  we show the  \NII$\lambda$6584/\Ha\ vs. \OIII$\lambda$5007/\Hb\  resolved diagram (hereafter \NII-BPT diagram).  We use the classification  of  \citet[][hereafter K03]{Kauf03} to select star forming (SF) regions (solid black line in Fig. \ref{fig:BPTMap}),  and the criteria of \citet{Kewley06} to separate composite (regions between the solid and dashed demarcation in Fig.~\ref{fig:BPTMap}), and shocks or AGN-excited gas regions (right of the dashed line in Fig. \ref{fig:BPTMap}). { From all the spaxels in the galaxy,} we find that 97.8$\%$ correspond to SF-like regions, 1.8 $\%$ to Composite-like regions, and 0.4$\%$ of AGN-like regions. 

In Fig.~\ref{fig:BPTMap} (right), we show the diagram \SII$\lambda$6717,31/\Ha\ vs. \OIII$\lambda$5007/\Hb\,  (hereafter \SII-BPT diagram). We use the demarcations of \citet{Kewley06}. Similarly, we find that 85$\%$ of spaxels correspond to SF-like regions, 7.7$\%$ to low ionization emission regions \citep[LIERs,][]{Belfiore16}, and 7.3$\%$ to Seyfert-type regions.  Error bars in Figures \ref{fig:BPTMap} and \ref{fig:MetGrad} of this paper are estimated using the `Propagate' package in `R'.




The position of ULX-1 lies in between 2 fibers, hereafter ULX-1(\textbullet) and ULX-1($\ast$). We find that ULX-2, has very low emission in  \NII$\lambda$6584 and \Hb\  (see the Appendix), for which we were unable to locate it on the BPT diagrams. As for ULX-3, a single fiber coincides with its location.  The GMS spectra of the fibers closest to the ULXs are shown in the Appendix. Both, ULX-1 and ULX-3 are located in the SF region of the two BPT diagrams analyzed  (see Fig. \ref{fig:BPTMap}). The location of  ULX-1(\textbullet) and ULX-1($\ast$) in the BPT diagrams are discrepant, despite being adjacent to each other in the galaxy. Part of this disagreement may be due to the low S/N of the ULX-1(\textbullet) spectra (see the Appendix), and/or to contamination from nearby sources affecting each spaxel differently.

Based on Fig.~\ref{fig:BPTMap}, the spectra of ULX-1 and ULX-3 are dominated by stellar photoionization, allowing us to use strong line diagnostics based on nebular photoionization as described in the next section.

\section[]{Metallicity and Ionization parameter} \label{MetIonGrad}


The `direct method' to estimate the oxygen abundances, or the electron-temperature T$_{\rm e}$ method, relies on measurements of auroral to nebular emission lines  \citep[e.g., \OIII $\lambda$$\lambda$5007,4959/\OIII $\lambda$4363,][]{Pilyugin05}. Since auroral lines are usually weak, and in our case, outside the spectroscopic range of the red setup of the GMS spectra, the gas metallicities are estimated using strong emission lines \citep{Pagel79,Alloin79}. From the numerous strong-line calibrations available, it is been shown that the differences between  diagnostics based on strong emission lines correlate strongly with the ionization parameter \citep[e.g.,][]{Ho15}. Hence, for our analysis we selected two independent methods, described below, that account for the ionization parameter.


\subsection[]{Gas metallicities from the $S$-calibration} \label{Scalibration}

The first method we use is the $S$-calibration \citep{Pil06}, which relies the following strong line ratios: S$_2$ = {\SII} ${\rm \lambda 6717 + \lambda 6731}$ / ${\rm{H\beta}}$, R$_3$ =  {\OIII} ${\rm{\lambda 4959 + \lambda 5007}}$ / ${\rm{H\beta}}$, and  N$_2$ =  {\NII} ${\rm \lambda 6548 + \lambda 6584}$ / ${\rm{H\beta}}$. For a detailed description of the method, see \citet{Pil06}. The $S$-calibration takes into account dependencies on the ionization parameter, is compatible with the metallicity scale of {\HII}  regions, and shows a smaller scatter compared to auroral line metallicities.

\begin{figure*}
\centering
{\includegraphics[width=0.45\textwidth]{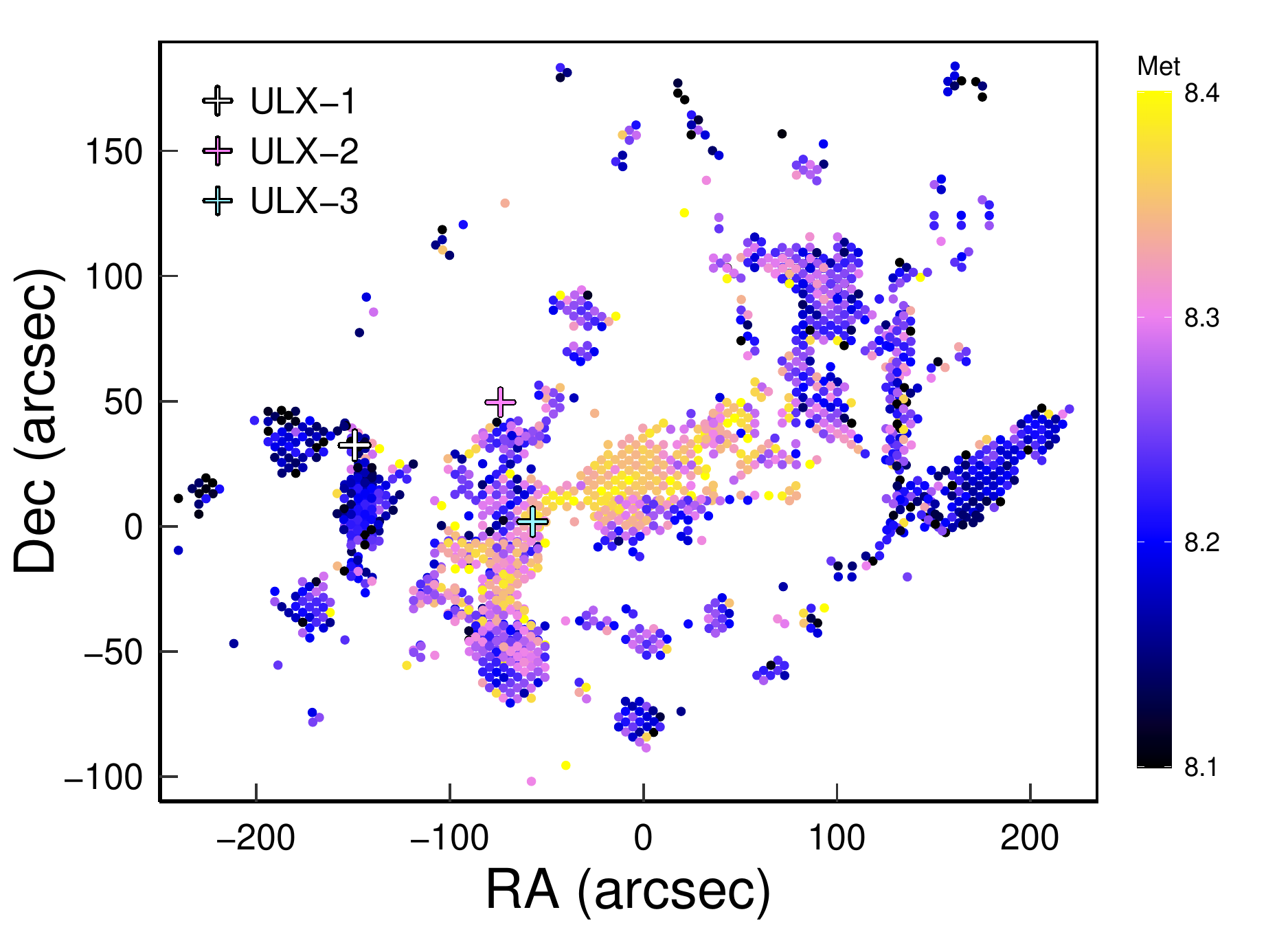}}
{\includegraphics[width=0.45\textwidth]{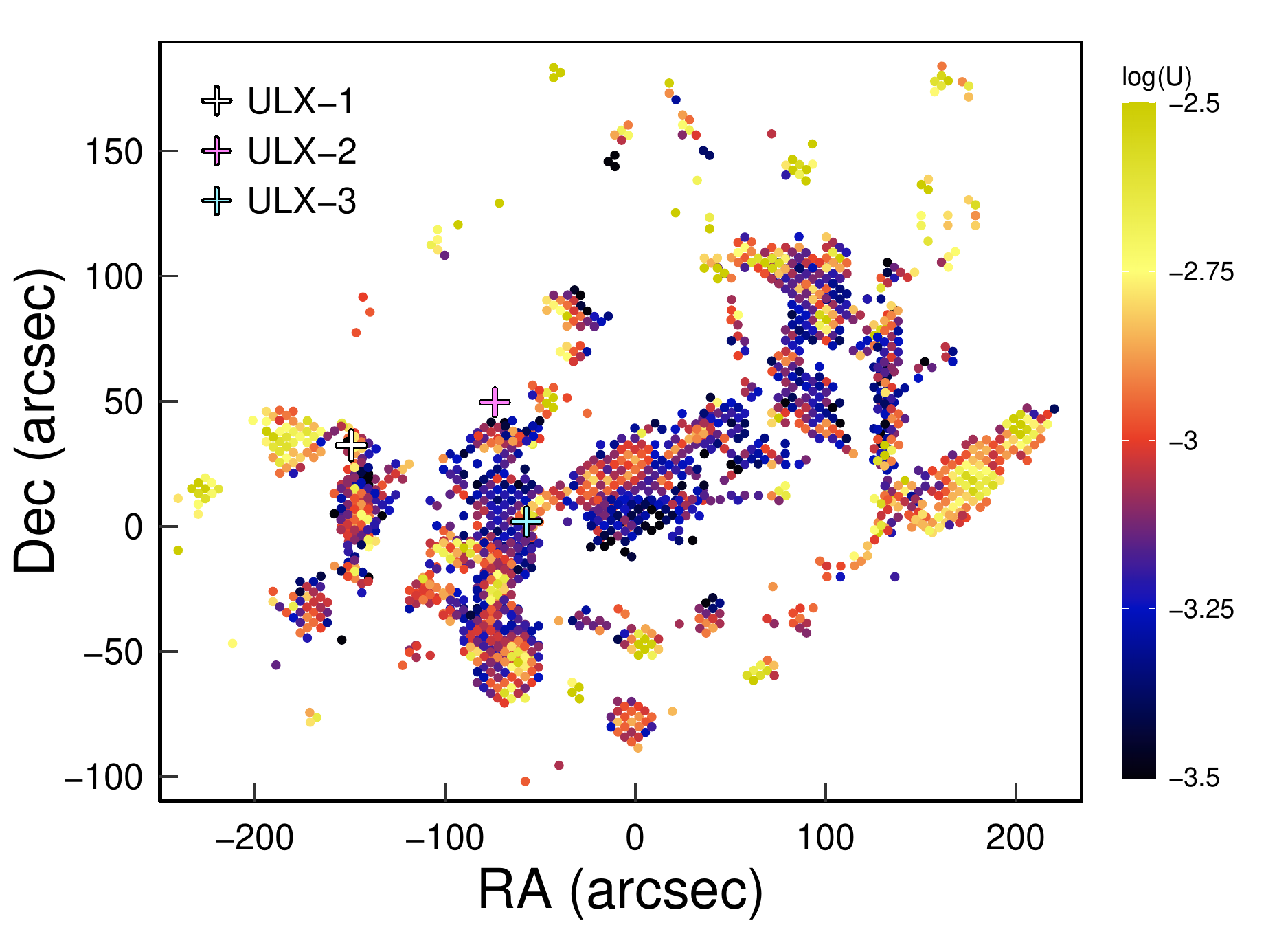}}
\caption{Metallicity and Ionization maps of NGC 925 color coded as indicated in the right bar. The white, pink and cyan crosses indicate the positions of ULX-1, ULX-2 and ULX-3, respectively.  \label{fig:MetMap}}
\end{figure*}

A signal-to-noise (S/N) cut of 3 is applied in each of the emission lines used in the method. Additionally, we select only SF regions using the standard  \NII-BPT diagram \citep{Baldwin81}, adopting the discrimination of K03, as shown in Fig. \ref{fig:BPTMap}. Our final SF sample consists of 1692 spaxels. The metallicity map for NGC 925 is shown in Fig. \ref{fig:MetMap} (left panel).

In order to further characterize the ULXs environment, we also estimate the ionization parameter U,  defined as the ratio of ionizing photon density to hydrogen density. The ionization parameter has proven to be a useful diagnostic tool, since any force that clears ionized gas from the interior of an HII region will have the effect of reducing the ionization parameter \citep{Yeh12}.


We estimate the ionization parameter following the prescription of  \citet{Dors17}:

\begin{equation}\label{IonDors}
{ {\rm log(U)_D}}  = c \times S2 + d
\end{equation}

where S2 = log(\SII  $\lambda \lambda$  6717,31 / \Ha), $c = -0.26 \times (Z/Z_{\sun}) - 1.54$, and $d= -3.69 \times (Z/Z_{\sun})^2 + 5.11 \times (Z/Z_{\sun})$ -5.26. { We used the oxygen abundance of the Sun as Z$_{\sun}$=8.69 \citep{Amarsi18}.}
The resulting ionization map is shown in Fig. \ref{fig:MetMap} (right panel). 


To estimate the metallicity and ionization gradients, we use the isophotal radius R$_{25} = 321.6$~arcsec from \citet{Schmidt16}. We estimate the inclination $i$ = 68.7$^{\circ}$ and position of the major axis (PA)=293.3$^{\circ}$  through the rotational curve using the approach of \citet{Pilyugin19}. Linear fits are performed in ``R" with the package ``{\textsc HYPER-FIT}" see \citet{Robotham15} for details. 

\begin{figure}
{\includegraphics[width=0.5\textwidth]{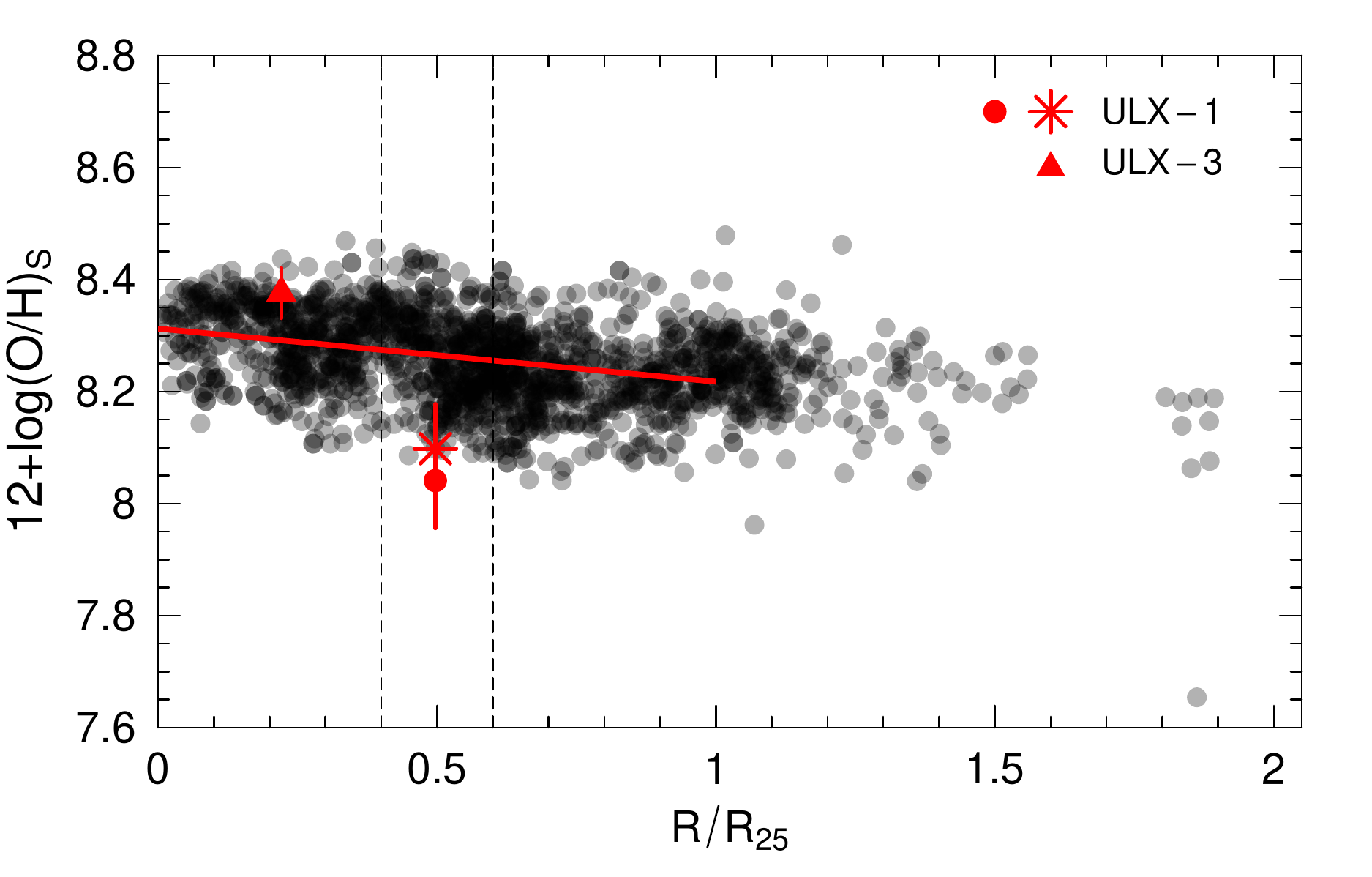}}
{\includegraphics[width=0.5\textwidth]{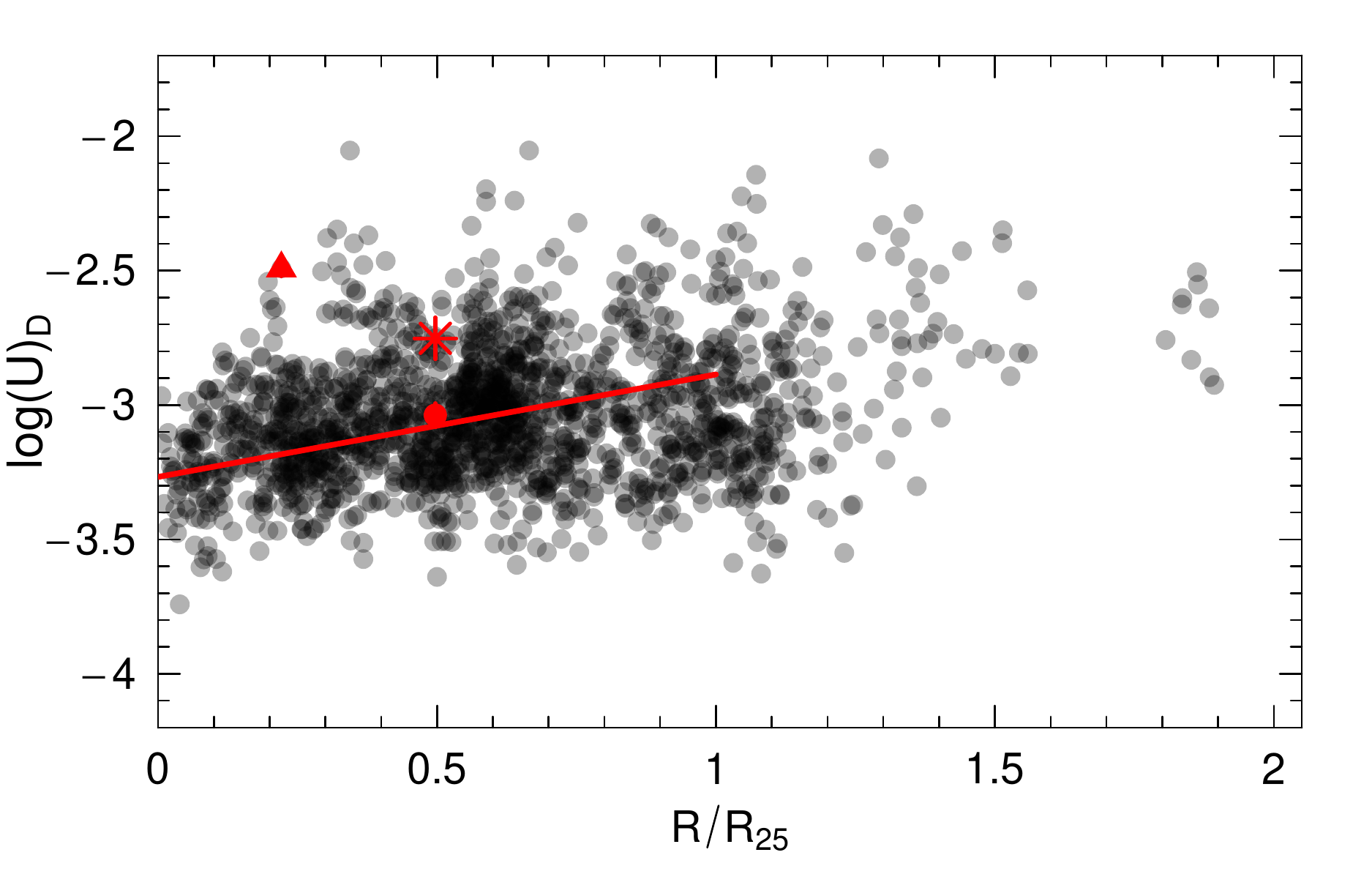}}
\caption{Metallicity and ionization gradient for  NGC 925. Data from the 2 fibers closest to the ULX-1 position are shown with a red asterisk and circle, while ULX-3 is shown with a triangle. \label{fig:MetGrad}}
\end{figure}

The metallicity gradient, using galactrocentric distances R$_{\rm g}$=R/R$_{25}$, is shown in Fig. \ref{fig:MetGrad}. The fitted parameters yield:

\begin{equation}\label{MetGradFit}
{\rm 12+log(O/H)_S} = - 0.0948(\pm0.0056) {\rm R_g} + 8.3124 (\pm 0.0038)
\end{equation} 

with a root mean square error (RMSE) of the residuals of 0.07 dex. We define the  RMSE as $\sqrt{\sum_{i=1}^{n} (\hat{y}_i-y_i)^2 / n }$, where $\hat{y}_i$ are the predicted values, and $y_i$ the observed values. The metallicity gradient we find is typical for galaxies of this mass \citep[e.g.,][]{Kreckel19, Igor19a}.



Similarly, we analyze the ionization gradient, obtaining the following:

\begin{equation}\label{IonGradFit}
{\rm log(U)_D}= 0.3813 (\pm0.0248) {\rm R_g} - 3.267 ( \pm0.016)
\end{equation} 

with a RMSE = 0.2427 (see Fig. \ref{fig:MetGrad}). The values of metallicity and ionization for ULX-1 and ULX-3 are listed in Table \ref{table:ComMet}.



\begin{table*}
\caption{\label{table:ComMet}  { Summary of ULXs properties. From left to right, gas metallicities given by the $S$-calibration, ionization parameter  from \citet{Dors17};  from BEAGLE: gas metallicities, ionization parameter at the edge of the Str\"{o}mgren sphere (see text), and interstellar metallicity.}}
\centering
 \begin{tabular}{l c  c c c c} 
 \hline
    & 12+log(O/H)$_{\rm S}$  & log(U)$_{\rm D}$ & 12+log(O/H)$_{\rm B}$ & log(U$_{\rm S}$)$_{\rm B}$ & (Z$_{\rm ISM}$)$_{\rm B}$   \\ [0.5ex] 
 \hline 
  ULX-1 (\textbullet) &  8.04$\pm${ 0.17} & -3.03 $\pm$ {0.09}  & 8.26 {$\pm$ 0.02} & -3.13 {$\pm$ 0.02} & 0.0070 {$\pm$ 0.0003}\\
 ULX-1 ($\ast$) & 8.10 $\pm$ {0.16} & -2.75 $\pm$ {0.09 } & 8.22 {$\pm$ 0.01} & -2.99 {$\pm$ 0.01} & 0.0064 {$\pm$ 0.0002}\\ 
 ULX-3 ($\blacktriangle$)& 8.37 $\pm$ {0.09}  & -2.50 $\pm$ {0.06 } & 8.7078 {$\pm$ 0.0001}& -3.1253 {$\pm$ 0.0005}  &  0.0134 {$\pm$ 0.0001}\\

 \hline
\end{tabular}
\end{table*}


\begin{figure*}
{\includegraphics[width=0.35\textwidth]{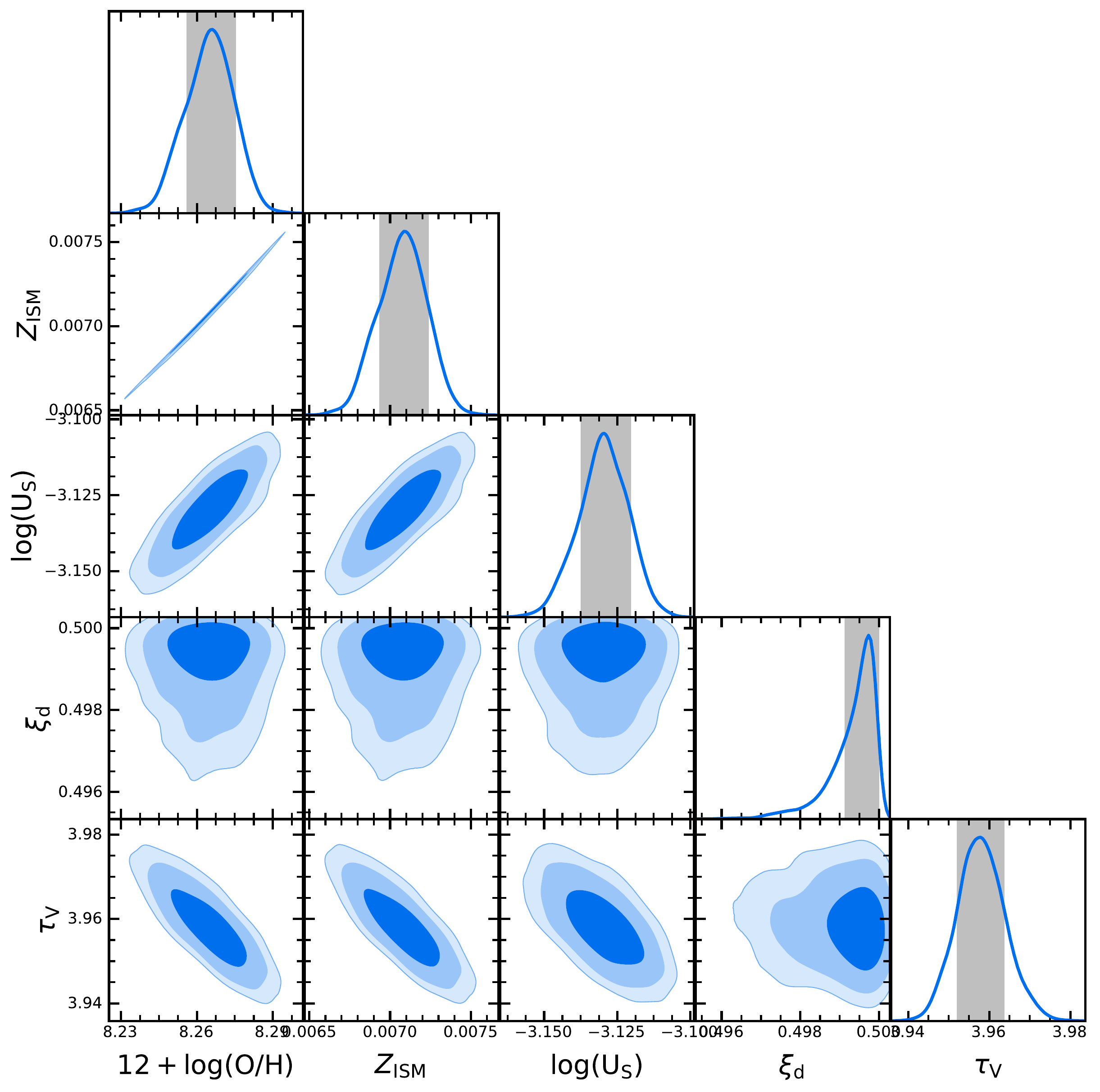}}
{\includegraphics[width=0.35\textwidth]{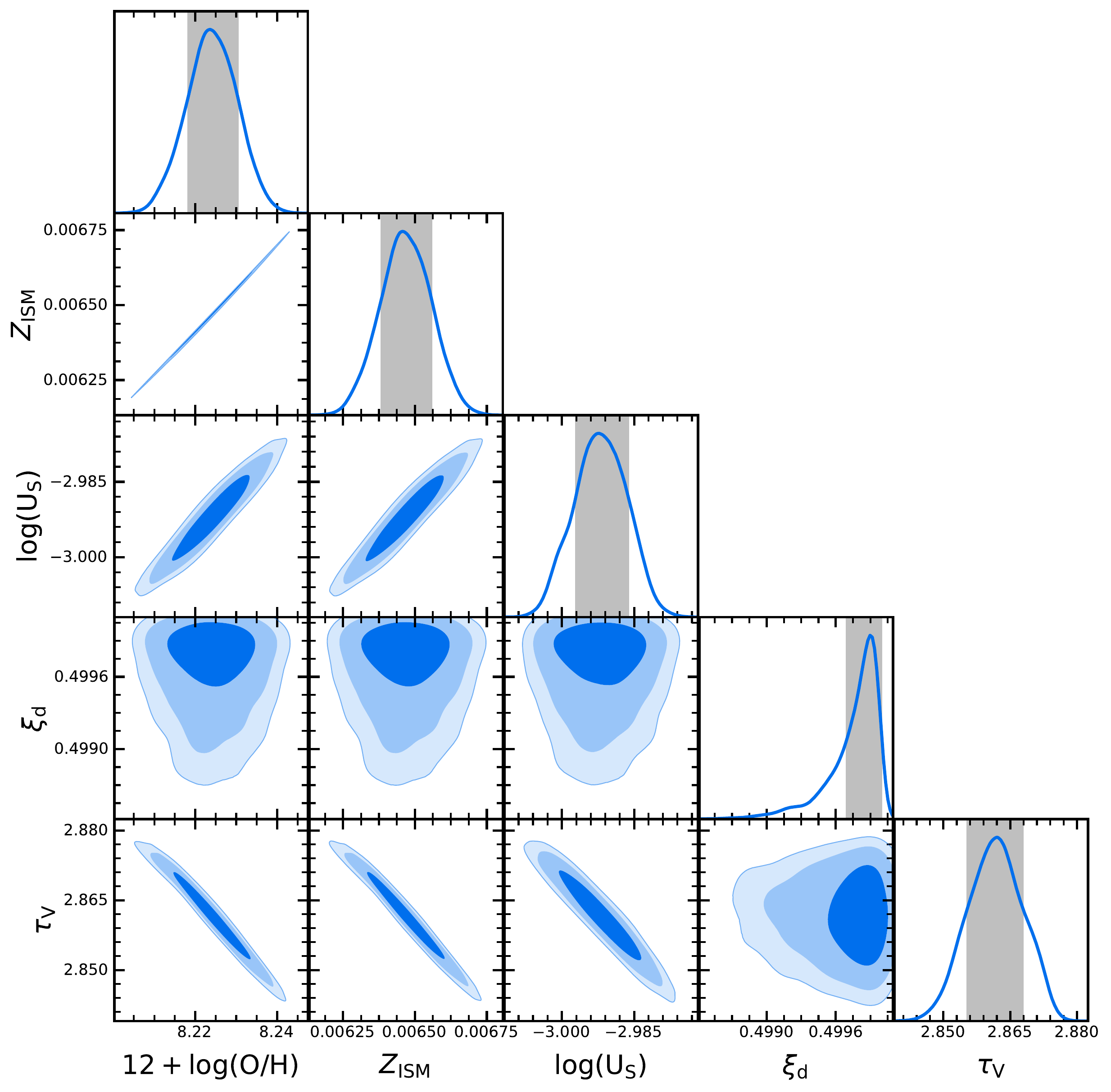}}
{\includegraphics[width=0.35\textwidth]{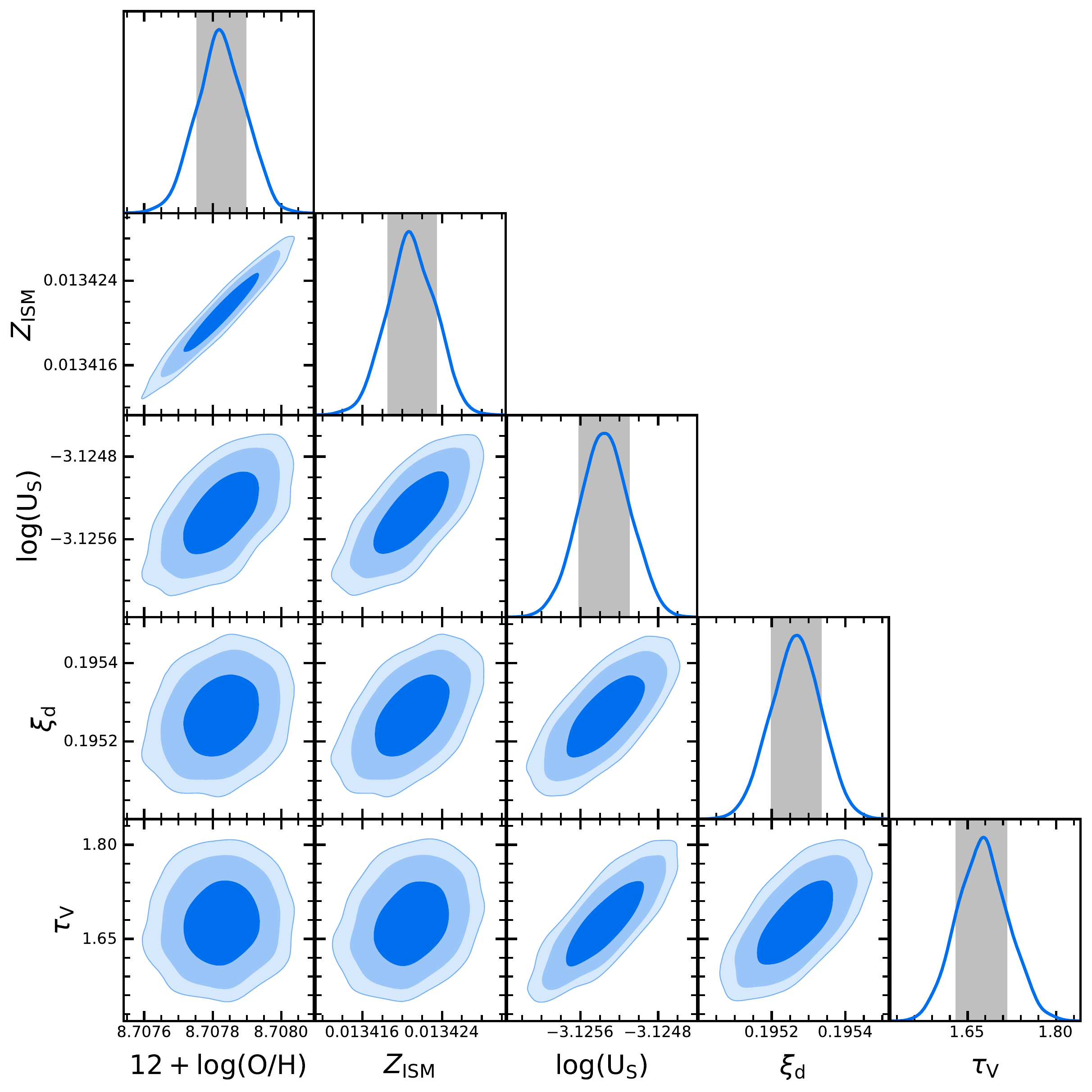}}
\caption{Two dimensional joint probability distributions and one dimensional marginal posterior probability 
distributions of the physical properties inferred by fitting the optical line fluxes with \textsc{beagle}. Panels from left to right correspond to ULX-1(\textbullet), ULX-1($\ast$) and ULX-3($\blacktriangle$). 
From left to right (bottom to top): oxygen abundance 12+log(O/H), interstellar metallicity Z$_{\rm ISM}$, ionization parameter, 
log(U$_{\rm S}$) and the dust-to-metal mass ratio, $\xi_{\rm d}$, and the optical depth in the V band, $\tau_{\rm V}$. 
The shaded blue contours indicate one, two, and three standard deviations from the median values, 
while the shaded gray rectangles indicate one standard deviation from the median values (68$\%$ credible interval).} \label{trianglePlot}
\end{figure*}\vspace{1cm}




Interestingly, Fig.~\ref{fig:MetGrad} indicates that the data from the two fibers overlapping with ULX-1 have a very low gas metallicity for their galactocentric distance. Furthermore, ULX-3 exhibits a very high ionization parameter. This is discussed further in \S \ref{Sect:Discussion}.

\subsection[]{Gas metallicities from BEAGLE} \label{BEAGLEMet}

To confirm the previous result using a second independent method, we use  the Bayesian spectral analysis tool \textsc{beagle} \citep{Chevallard2016}, 
which incorporates models for the emission from stars and photoionized gas.
Specifically, the stellar models \citep[latest version of the][stellar evolutionary code]{Bruzual03} are computed using a 
standard \cite{Chabrier2003} IMF with an upper mass cutoff of 100 M$_{\odot}$. These models have been combined with \texttt{CLOUDY} 
\citep[v13.03][]{Ferland2013} to compute the corresponding nebular emission from the ionized gas in HII regions \citep{Gutkin16}.
For our analysis, we adopt a constant star formation history and the model of \cite{Charlot2000} to describe dust attenuation. 
One interesting feature of BEAGLE is that it takes into account multiple parameters that can impact the line ratios simultaneously and thus the reported error bars appropriately reflect the parameter degeneracies. We re-measured the flux calibrated spectra using the software Pyspeclines\footnote{https://pypi.org/project/pyspeclines/} by fitting Gaussians to the spectra prior the continuum subtraction and correction for stellar absorption. This is because BEAGLE templates include both continuum and stellar absorption features.
The emission lines fed into BEAGLE are \Ha, \Hb, [OIII]$\lambda$5007, [NII]$\lambda$6584, [SII]$\lambda$6716, 6731.
We use BEAGLE to fit the line fluxes of the three ULX spectra, as well as the 399 spaxels at the galactocentric distances  0.4 $<$ R/R$_{25}$ $<$ 0.6 indicated in Fig.~\ref{fig:MetGrad}.


Our results are illustrated in Fig.~\ref{trianglePlot}, which shows the probability distributions of the nebular physical 
properties: oxygen abundance 12+log(O/H)$_{\rm B}$, interstellar metallicity (Z$_{\rm ISM}$)$_{\rm B}$, the dust-to-metal mass ratio ($\xi_{\rm d}$)$_{\rm B}$, and  the ionization parameter at the edge of the Str\"{o}mgren sphere log(U$_{\rm S}$)$_{\rm B}$.  
We note that the two ionization parameters inferred with the BEAGLE and \citeauthor{Dors17} methods are not directly comparable; the first is the ionization parameter at the edge of the Str\"{o}mgren sphere, while the second at the inner edge of the gas cloud.
A summary of the median values for 12+log(O/H)$_{\rm B}$, log(U$_{\rm S}$)$_{\rm B}$ and (Z$_{\rm ISM}$)$_{\rm B}$ is given in Table \ref{table:ComMet}.  We note that the gas metallicities given by BEAGLE depend on a grid with values of (Z$_{\rm ISM}$)$_{\rm B}$ and ($\xi_{\rm d}$)$_{\rm B}$ \citep{Gutkin16}, which would add to the gas metallicities an intrinsic error of $\sim$0.05 dex based on the average spacing of the grid.

We find that the difference in gas metallicities between the $S$-calibration and BEAGLE ranges from $\sim$0.1 to $\sim$0.4 dex. This is in agreement with previous work comparing photoionization-based metallicities with the  direct-Te method \citep[e.g.,][]{Vale16}.
Clearly, the gas metallicity 12+log(O/H) in the environment of ULX-1 is lower in comparison to ULX-3. On the other hand, the ionization parameter of ULX-3 is just slightly higher than the ULX-1 spectra.

\begin{figure}[h!]
\centering
{\includegraphics[width=0.4\textwidth]{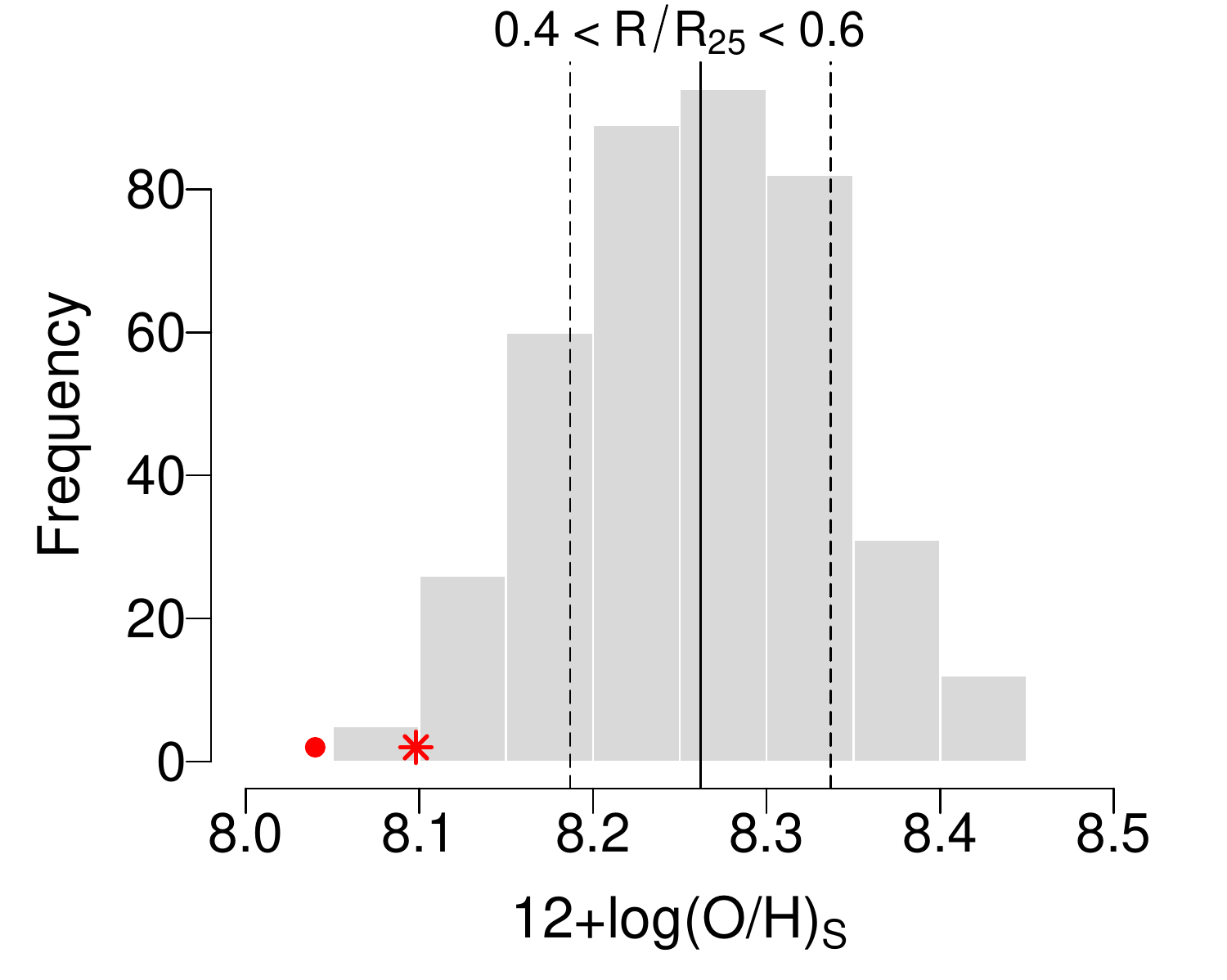}}
{\includegraphics[width=0.4\textwidth]{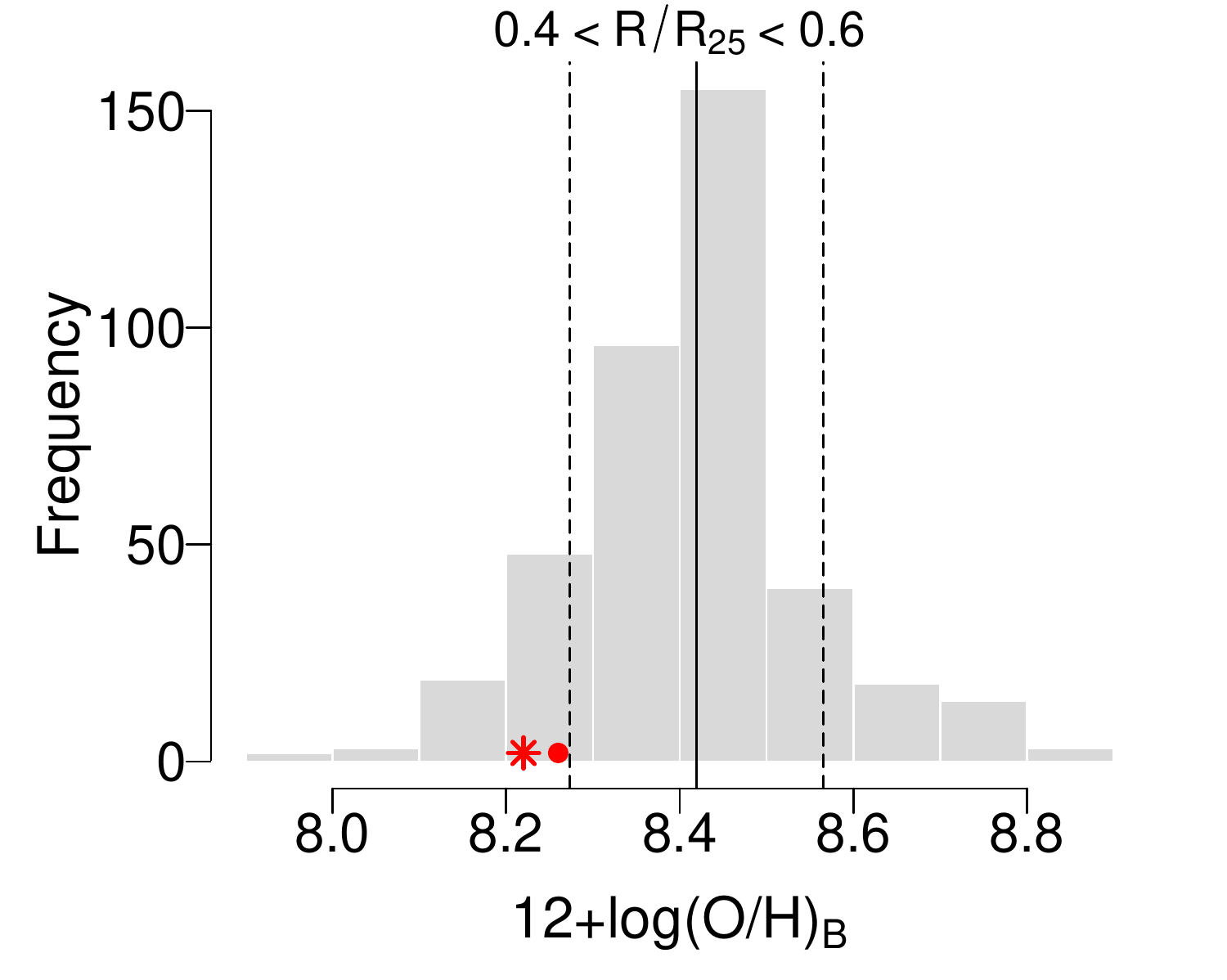}}
\caption{Histogram of the metallicity for spaxels in the range 0.4 $<$ R/R$_{25}$ $<$ 0.6. The median and 1-$\sigma$ scatter of the sample are shown by the solid and dashed black lines, respectively. Data of the two fibers closer to the ULX-1 position are shown with an asterisk and circle. The top and bottom histogram show metallicities estimated by the $S$-calibration, and BEAGLE, respectively. \label{fig:HistULX}}
\end{figure}

In order to highlight the low metallicity values of the ULX-1 region using the two different metallicity methods, we show a histogram of the 399 spaxels around  ULX-1 in Fig.~\ref{fig:HistULX}, in the interval 0.4 $<$ R/R$_{25}$ $<$ 0.6. The median metallicity using the $S$-calibration in this interval is 8.26, with a 1-$\sigma$ scatter of 0.075. The gas metallicities of ULX-1($\ast$) and ULX-1(\textbullet) are 0.16 and 0.22 dex below the median value, respectively.
Since comparisons between different metallicity methods reveal large discrepancies \citep[e.g.,][]{Kewley08, LopezSanchez12}, we also show { in Fig.~\ref{fig:HistULX} the gas metallicity distribution estimated by BEAGLE for the same spaxels}. {  The median metallicity  is 8.42, with a 1-$\sigma$ scatter of 0.14, where the gas metallicities of ULX-1($\ast$) and ULX-1(\textbullet) are 0.12 and 0.16 dex below the median value, respectively. }
Both the $S$-calibration and BEAGLE estimations confirm that the gas metallicities of ULX-1($\ast$) and ULX-1(\textbullet) are very low even  compared to other HII regions in the galaxy at the same galactocentric radius.

\section[]{Discussion} \label{Sect:Discussion}



\subsection{Summary of observational constraints}

We find that ULX-1 and ULX-2 are associated with nebulae of the order of (major/minor axis) 285/285 and 285/232 pc, respectively. The dimensions of these nebulae  suggest they are ULX bubbles, which can be distinguished from SNRs and HII regions due to their larger size  \citep[$\sim$100-300 pc,][]{Pakull08}. ULX-3  is  associated with two nearby H$\alpha$ blobs that are separated by $\sim$350 pc. Furthermore, the high velocity dispersion of the nebulae and blobs suggest the presence of shocks, although we cannot resolve any expansion signature with the PUMA data.  On the other hand, the analyzed BPT diagrams indicate the ionization is consistent with HII regions. We explore as well
the [SII]/\Ha\ ratio, that can distinguish between shock-ionized and photoionized regions, where shock-ionized regions have [SII]/\Ha\ $>$ 0.4 \citep{Dodorico80}. For our data, these ratios are 0.43, 0.29, and 0.2 for ULX-1(\textbullet), ULX-1($\ast$) and ULX-3, respectively. Therefore, while it indicates that ULX-1(\textbullet) is possibly consistent with a shock-ionized region, the result is inconclusive as ULX-1($\ast$) is below this threshold.

To further explore the possibility of shocks in the ULX-1 spaxels and their impact on the gas metallicity, we use the  MANGA sample of \citet{Pilyugin20}. From this sample, we examine the gas metallicity, using the S-calibration, of SF-type regions with gas velocity dispersions of $\sim$100 km/s, and spaxels with sizes of $\sim$300 pc. We find that the gas metallicity for these regions follows the metallicity gradient of the host galaxy. Therefore we conclude that if shocks were present, they would have a minimum impact on the gas metallicities.


We note that the combination of different resolutions and depths make the maps  from PUMA and the GMS not directly comparable. For instance, the high resolution data of PUMA does detect \Ha\  emission in the vicinity of ULX-2 (see Fig. \ref{fig:PUMA}). However, the same region is not as bright in the \Ha\ surface brightness map from the GMS (Fig. \ref{fig:HaMap}). This can be explained due to the different resolution, plus the extra requirement of the \Hb\  line in emission to correct for the extinction in the GMS data.  Additionally, data from PUMA show two bright blobs around the location of ULX-3. It is likely that the lower spatial resolution of GMS is mixing the information of these blobs, and that a deeper image from PUMA could potentially unveil more gas around the ULX-3 position.

The ULX emission line ratios are located in the star-forming region of the BPT diagram, thus allowing us to use conventional prescriptions for gas metallicity and ionization. 
We note however, that photoionization models predict that AGN-ionized regions with low metallicity can populate the left side of the [NII]-BPT diagram. For instance, if the emission lines of any of the ULXs are powered by accretion around a black hole, we would not identify it from the  BPT diagram if it is in a regime with Z $<$ $\sim$  1/3 Z$_{\sun}$ \citep[e.g.,][]{Groves06,Feltre16}, which is the case for ULX-1. 

From the GMS observations, we find that the two  fibers closest to ULX-1 have a very low metallicity with respect to their galactocentric distance. This result is supported by metallicities estimated with the $S$-calibration and BEAGLE (see Fig.~\ref{fig:MetGrad} \& \ref{fig:HistULX}). On the other hand, their ionization parameter seems to be within the dispersion. 

Overall, the results for ULX-1 agree with \citet[][]{Heida16}, who suggest a bright nebulae in the vicinity of ULX-1 with very low gas metallicity. However, we are uncertain on the origin of the emission lines, as pointed out earlier in this section. 

We emphasise that IFU analyses, with surveys such as Metal-THINGS, enable the study of the entire region surrounding ULXs, and not just the potentially contaminated regions very close to ULXs, as is the case for single fiber or multi-object spectroscopy (MOS) of point-like sources.


\subsection{Physical nature of the ULXs}

Results in the literature show that ULXs are frequently found in low metallicity environments \citep[e.g., ][]{Linden10, Mapelli10, Prestwich13, Lopez19, Basu16}.  Amongst the possible explanations, \citet{Linden10} show that the majority of the ULX population can be explained by the tail of the high-mass X-ray binary (HMXB) population, where the low-metallicity environment favours active Roche lobe overflow (RLO-HMXBs). In this scenario, the lower metallicity binaries will have, on average, tighter orbits which will result in an increased number of RLO-HMXBs that can drive much higher accretion rates. Therefore, lower metallicity HMXB populations are expected to be more luminous  \citep[e.g., ][]{Linden10,Fragos13}.



A different ULX origin is proposed by \citet{Kim17}, who conclude that a ULX in NGC 5252 is most likely the nucleus of a small, low-mass galaxy accreted by NGC 5252. To explore this possibility, we report a histogram of the metallicity of the 27 spaxels within a radius of $\sim$400 pc centred on ULX-1, and find a very similar distribution as shown in Fig.~\ref{fig:HistULX} using the $S$-calibration,  with a median metallicity of 8.24, and a 1-$\sigma$ scatter of 0.076. Since the low metallicity gas is not spatially extended, we consider this possibility highly unlikely for ULX-1.

In contrast to ULX-1, the gas metallicity and ionization parameter of ULX-3 are both high (Fig.~\ref{fig:MetGrad}).  This is more consistent with the presence of a neutron star within an evolved stellar population, as postulated by \citet{Earnshaw20}. This scenario is not uncommon, since several neutron star ULXs have been discovered in recent years \citep[e.g.,][]{Bachetti14,Furst16,Israel16,Carpano18}.





\section{Conclusions}\label{Conclusion}

Using data from the Metal-THINGS survey, we present an optical analysis of three ULX sources located in NGC 925. We use two complementary sets of observations: high resolution \Ha\  data from the Fabry-Perot interferometer PUMA, and  IFU spectroscopy from the GMS. While the spatial resolution of PUMA is $\sim$3 orders of magnitude higher than the GMS, the spectral range of the GMS allows us to measure several strong emission lines and derive properties such as the gas metallicity and ionization. Our main conclusions are:



\begin{itemize}[leftmargin=*]



\item[--]  With the PUMA observations, we determined the velocity of the ionized gas around the three ULXs and found the size of the associated nebula,  which range in size from 174 to 285 pc.  The large dimensions of ULX-1 and ULX-2 are consistent with ULX-bubbles, while ULX-3 is associated with two blobs. The high velocity dispersion associated with the  nebulae and blobs suggest the presence of shocks ionizing the gas. The GMS data suggest a shock-ionized region for ULX-1(\textbullet) since [SII]/\Ha\ $>$ 0.4, however, since ULX-1($\ast$) is below this threshold, and both ULX spaxels are located in the SF region of the BPT diagrams, this result is inconclusive.


\item[--] Resolved BPT diagrams were explored for NGC 925. We find that both, ULX-1 and ULX-3 are located in the star-forming area of the \NII-BPT and \SII-BPT diagrams.   However, we note that AGN-ionized regions of low metallicity  can populate the star-forming region of the BPT diagrams, making them difficult to identify. The spectra corresponding to ULX-2 have low S/N, and thus we were unable to measure any properties.

\item[--] We estimated gas metallicities using the $S$-calibration \citep{Pil06}, and  ionization using the prescription of \citet{Dors17}. A negative metallicity gradient is found using galactrocentric distances R/R$_{25}$, consistent with previous results for spiral galaxies.  In contrast, the ionization parameter seems to increase radially throughout the galaxy, although with a large dispersion. 

\item[--] { The spectra of the three ULXs and all spaxels in a galactocentric distance range around ULX-1 (0.4$<$R/R$_{25}$$<$0.6) were fitted using the Bayesian spectral analysis tool \textsc{BEAGLE}. Both the S-calibration and \textsc{BEAGLE} indicate that the two ULX-1 spectra are in an environment with a very low metallicity, even when compared with HII regions at the same galactocentric distance. }



\item[--] Our results from PUMA and the GMS agree with  \citet[][]{Heida16}, suggesting a bright nebula in the vicinity of ULX-1. Furthermore, the low gas  metallicity detected is also in agreement with several authors \citep[e.g., ][]{Mapelli10,Linden10,Fragos13}, who suggest that the { frequency of ULX occurrence is anti-correlated with metallicity. Within this scenario, the ULX population can be explained by the tail of the HMXB population, where the low-metallicity environment favours active Roche lobe overflow that can drive much higher accretion rates \citep{Linden10}.}

\item[--] In the case of ULX-3, we measure a high gas metallicity and ionization. This result would agree with accretion onto a neutron star located in an evolved stellar population region, as suggested by \citet{Earnshaw20}.
\end{itemize}

\acknowledgments
We acknowledge the careful examination by the referee.
This paper is based on observations with the 2.7m Harlan J. Smith Telescope, operated by The McDonald Observatory of The University of Texas at Austin; and   observations with the Observatorio Astron\'mico Nacional on the Sierra San Pedro M\'artir (OAN-SPM), Baja California, M\'exico. 
We  acknowledge David Doss and the staff at McDonald Observatory for their invaluable help during the observations. We thank the daytime and night support staff at the OAN-SPM for facilitating and helping obtain our observations.
We appreciate the insightful help of Michaela Hirschmann and Jorge Moreno.
M.A.L.L acknowledges support from the Carlsberg Foundation via a Semper Ardens grant (CF15-0384).
I.A.Z acknowledges support by the grant for young scientist's research laboratories of the National Academy of Sciences of Ukraine.
 L.S.P acknowledges support within the framework of the program of the NAS of Ukraine ``Support for the development of priority fields of scientific research'' (CPCEL 6541230).
 M.L.P.G acknowledges the support from European Union\textsc{\char13}s Horizon 2020 research and innovation programme under the Marie Sklodowska-Curie grant agreement No 707693.
 A.F. acknowledges the support from grant PRIN MIUR2017-20173ML3WW\texttt{\char`_}001.
 M.E.D.R is grateful to PICT-2015-3125 of ANPCyT (Argentina).
  S. D acknowledges support from the french ANR and the german DFG through the project "GENESIS" (ANR-16-CE92-0035-01/DFG1591/2-1).
  E.I \ acknowledges partial support from FONDECYT through grant N$^\circ$\,1171710.
I.F-C acknowledges support from IPN-SAPPI project 20201692.
 M.R acknowledges the UNAM-DGAPA-PAPIIT grant IN109919 and the CONACyT grant CY-253085. 
\\\\\vspace{20cm}

%

\facilities{2.7m Harlan J. Smith Telescope, McDonald Observatory, Texas; and  2.1m Telescope, OAN-SPM, M\'exico.}\\


\software{P3D (https://p3d.sourceforge.io);
		astropy \citep{astropy13};  
		IRAF \citep{IRAF1};
		``R" statistical programing language (http://www.R-project.org/);
		Pyspeclines (https://pypi.org/project/pyspeclines/);
		PySpecKit \citep{Ginsburg11};
          }\\\vspace{20cm}



\appendix \label{Apendix}

Spectra of the ULXs analyzed in this paper.

\begin{figure}[h!]
\centering
{\includegraphics[width=0.7\textwidth]{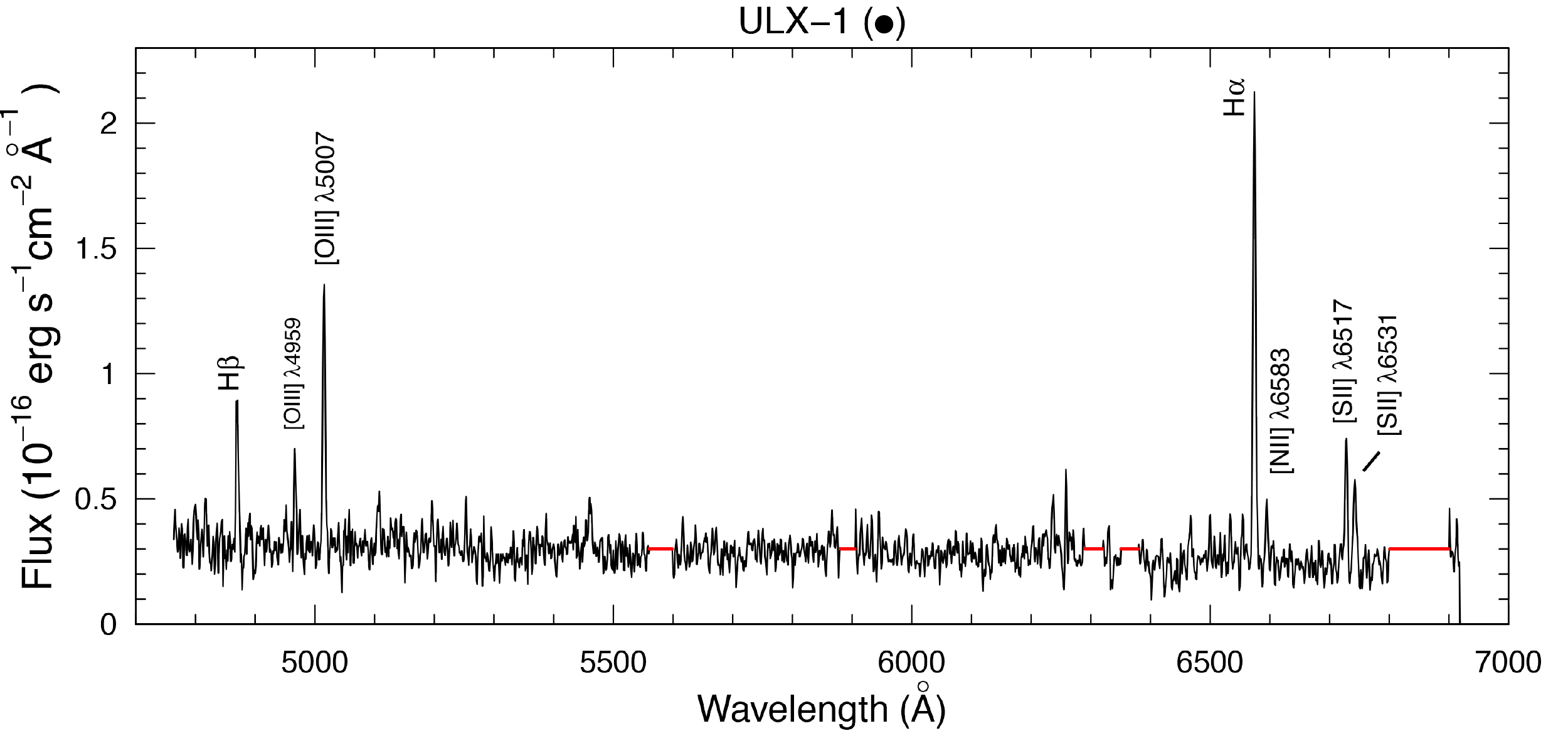}}
{\includegraphics[width=0.7\textwidth]{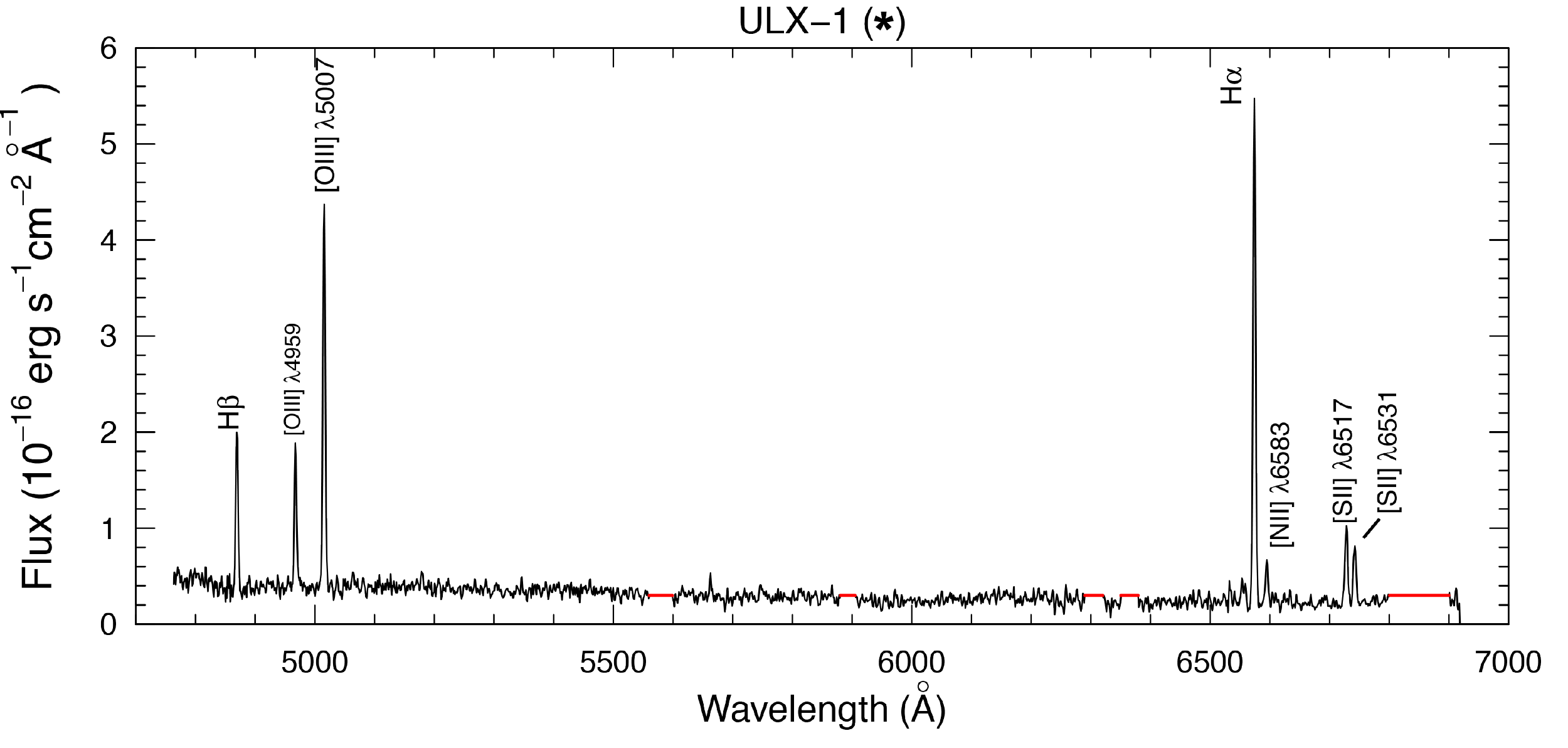}}
{\includegraphics[width=0.7\textwidth]{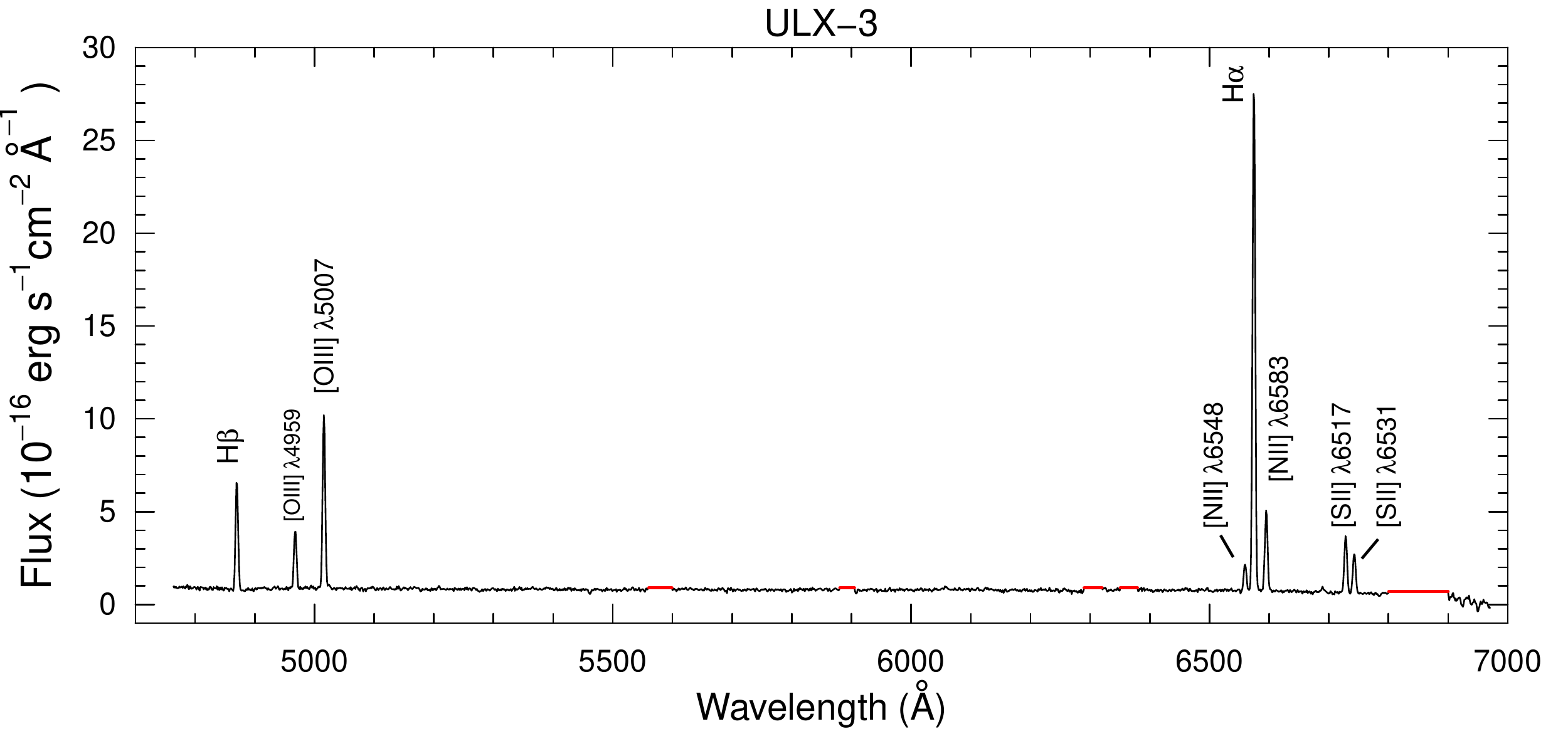}}
{\includegraphics[width=0.7\textwidth]{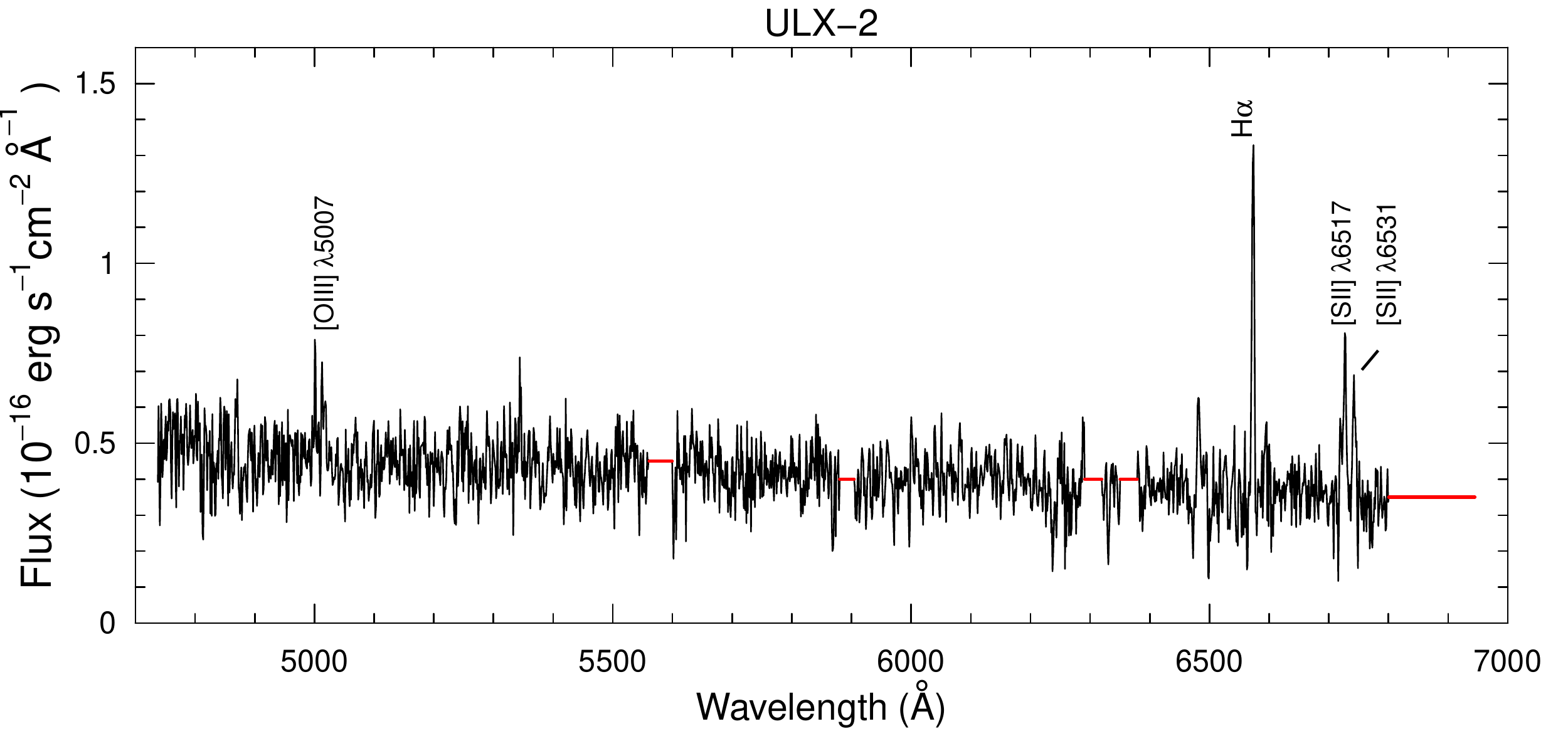}}
\caption{Spectra of the three ULX sources analyzed in this paper. Red horizontal lines indicate the masked parts of the spectra with strong sky emission lines. \label{fig:SpectraULX}}
\end{figure}

\bibliography{NGC925_ULXs_V4}{}
\bibliographystyle{aasjournal}



\end{document}